%% file: 2025_Strutz_Kiers_Curtis_DASdesign.tex
\documentclass[]{preprint}

\title{Single and Multi-Objective Optimization of Distributed Acoustic Sensing Cable Layouts for Geophysical Applications}
\author[1]{Dominik Strutz}
\author[2]{Tjeerd Kiers}
\author[1]{Andrew Curtis}

\affil[1]{School of Geosciences, University of Edinburgh, Edinburgh, UK}
\affil[2]{Institute of Geophysics, ETH Zurich, Zurich, Switzerland}

\submittedto{Geophysical Journal International}

\begin{document}

\maketitleabstract{
    \begin{summary}{Abstract}\label{sec:abstract}
            We present a systematic approach to optimise distributed acoustic sensing (DAS) fibre-optic cable layouts using global optimisation techniques. Our method represents cable geometries using splines, enabling efficient exploration of layouts while respecting physical deployment constraints. The use of evolutionary algorithms enables single and multi-objective optimisation, taking into account complex design constraints such as terrain, accessibility, exclusion zones and cable length, while allowing efficient parallelisation of the optimisation process. We demonstrate the approach on a real-world case study, optimising the layout of a DAS cable for monitoring slope stability in the Cuolm da Vi area of Switzerland. We adapt design criteria for seismic source location problems, and for ambient noise surface wave tomography, to account for the unique characteristics of DAS, such as directional sensitivity patterns. The results show significant potential for improvements in source location accuracy and surface wave tomographic resolution by optimising cable layouts, highlighting the potential of this approach for optimising DAS deployments in various geophysical applications.
    \end{summary}
} 

\input{content.tex}

\bibliography{bibliography.bib}

\appendix
\input{appendix.tex}

\end{document}

%% file: content.tex
\section{Introduction}

Distributed acoustic sensing (DAS) has expanded observational capabilities in various seismological applications, including subsurface tomography \citep[e.g.,][]{Daley2013-FieldMonitoring,Daley2016-FieldMBM}, reservoir monitoring \citep[e.g.,][]{Mateeva2014-DistributedVSP}, volcanology \citep[e.g.,][]{Klaasen2021-DistributedColumbia, Fichtner2022-Fiber-opticResonance, Jousset2022-FibreEvents}, glaciology \citep[e.g.,][]{Walter2020-DistributedTerrain,Hudson2021-DistributedAntarctica}, earthquake-induced ground motion analysis \citep[e.g.,][]{Spica2020-UrbanSeismology, Yang2022-Sub-kilometerSensing}, and seismic event detection and monitoring \citep[e.g.,][]{Lindsey2017-Fiber-opticObservations, Martin2017-SensitivityArrays, Thrastarson2021-DetectingSensing, Hudson2025-TowardsNetworks,Miao2025-AssessingJapan,Baird2025-OceanNetworks,Baba2024-SeismicJapan}.

Compared to traditional individual seismic sensors, DAS offers several advantages such as high spatial density, and the ability to deploy sensing elements over long distances without the need for power at the points where data is recorded. These benefits make DAS particularly suitable for the creation of dense arrays, in environmentally sensitive or seafloor deployments and in remote or hazardous environments. However, the sensitivity of DAS cables poses a challenge for optimal cable placement, as the channels exhibit more extreme, directionally anisotropic, sensitivity patterns to waves incident on the fibre than single-component geophones \citep{Kuvshinov2016-InteractionCables,Martin2017-SensitivityArrays}. This directional sensitivity can lead to suboptimal data acquisition if the cable layout is not designed carefully.

Despite many studies on experimental design for geophysical applications such as source location and tomography \citep[e.g.,][]{Rabinowitz1990-OptimalApproach,Steinberg1995-ConfiguringSources,Curtis1999-OptimalExamples,Curtis1999-OptimalSurveys,Curtis2004-TheoryProblems,Curtis2004-TheoryProblems-01,Curtis2004-DeterministicSurveys,Maurer1998-OptimizedExperiments,Maurer2010-RecentDesign,Bloem2020-Experimental,Krampe2021-OptimizedCosts,Strutz2024-VariationalReservoir,Callahan2025-AnalysisDesign} and advances in design optimisation methodology \citep[e.g.,][]{Foster2019-VariationalDesign,Kleinegesse2021-Gradient-basedBounds,Mercier2025-DesigningApproach}, there remains a lack of systematic approaches for optimising DAS cable layouts. The Kangerlussuaq algorithm proposed by \citet{Fichtner2022-SimpleSensing} is a notable exception, but it begins from only a single initial cable layout that is locally perturbed, cannot be readily extended to multiple design criteria, and their quality criterion does not account for the directional sensitivity of DAS cables. The work of both \citet{Nasholm2022-ArraySpace} and \citet{Luckie2024-PerformanceGeometry} focuses on the effect of cable directionality on the performance of a small number of chosen layouts, but does not provide a systematic approach to optimise the cable layout.

In this study, we present a systematic global optimisation approach to optimising DAS cable layouts for any number of design criteria while ensuring the avoidance of obstacles or precluded areas. We demonstrate the optimisation process using two design criteria: optimising the cable layout for (1) linearised surface wave tomography, and (2) probabilistic seismic source location, both of which are adapted to account for the directional sensitivity of DAS cables. We therefore cover two common seismological applications, but also the two main ways of approaching experimental design: computationally efficient linearised experimental design that can scale to large problems, and fully Bayesian experimental design that accounts for the full non-linearity of the problem, but comes at a higher computational cost. The real-world applicability of the proposed method is demonstrated by optimising a DAS cable layout for different scenarios on the Cuolm da Vi slope instability.

\section{Methods}

This work incorporates concepts from distributed acoustic sensing, seismology, linearised and fully nonlinear Bayesian experimental design, and global optimisation. We now introduce the relevant methodology and how it applies to the optimisation of DAS cable layouts. Given the potentially broad scope of each of these multiple topics, we keep the introductions relatively brief and refer to the literature for more detailed treatments.

\subsection{Distributed Acoustic Sensing}

Distributed acoustic sensing (DAS) transforms standard fibre-optic cables into dense arrays of seismic sensors by measuring strain rate along the cable \citep{Lindsey2021-Fiber-opticSeismology}. It operates by firing a laser pulse through one end of the cable and detecting temporally-varying phase shifts in laser light backscattered by Rayleigh scattering at fibre impurities, where the variations are assumed to be caused mainly by strain along the cable axis. These measurements are averaged over a user-defined gauge length, which represents the spatial resolution of the system and determines the length of fibre over which strain rate is averaged to obtain each recorded signal \citep{Ajo-Franklin2019-DistributedDetection, Lindsey2021-Fiber-opticSeismology}. The locations at which these averaged (``distributed'') strain rates are recorded are referred to as channels, and the distance between these channels is referred to as channel spacing.

For the purpose of this study, we assume that the fibre-optic cable on which DAS is performed consists of a sequence of channels, each with a location and orientation. We treat the recorded data as point measurements located at the centre of each channel. Channel locations are denoted by $\left\{ \mathbf{c}_i \right\}_{i=1}^{N_\text{cha}}$ and the orientation of the cable at channel $i$ is given by the unit vectors $\left\{ \mathbf{u}_i \right\}_{i=1}^{N_\text{cha}}$, where $N_\text{cha}$ is the number of channels. A cable layout is then represented by the set of channel locations and orientations $\mathcal{L} = \{ \mathbf{c}_i, \mathbf{u}_i \}_{i=1}^{N_\text{cha}}$.

The main design constraint for a DAS cable layout compared to a conventional seismic survey is that the channels follow a continuous path. To achieve this, we parameterise the cable path using knot locations $\left\{ \mathbf{k}_i \right\}_{i=1}^{N_\text{knot}}$, where $N_\text{knot}$ is the number of knots. We can use these knots to define a continuous path by interpolation between the knots with a spline function. In this study we use B-splines \citep{Virtanen2020-SciPyPython, DeBoor2001-PracticalSplines} of a given degree $k$, but other algorithms that construct a continuous path from a set of knots could be used equally well.

Once the spline has been constructed, we can interpolate the channel locations $\mathbf{c}_i$ along the spline at any desired (here, constant) channel spacing. The knot locations and channel spacing therefore define the number of channels $N_\text{cha}$. We can then calculate the channel orientations $\mathbf{u}_i$ by taking the 3D spatial derivative of the spline at the channel locations, making use of a digital elevation model (DEM) if one is available.

\begin{figure}
 \centering
 \includegraphics[width=0.7\textwidth]{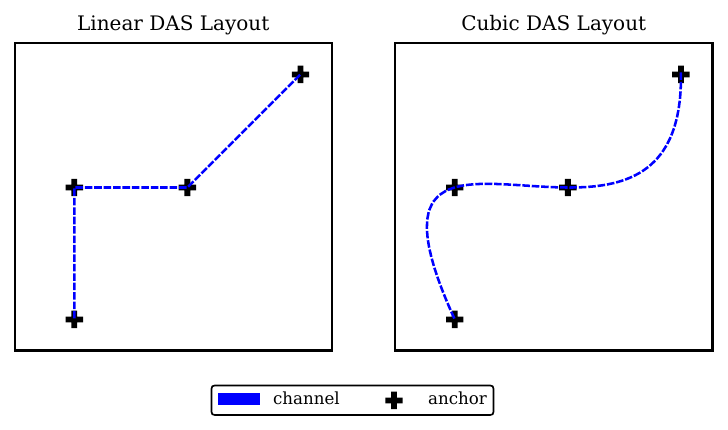}
 \caption{
  Example of a DAS cable parametrisation with similar knots but different spline types: linear (k=1) and cubic (k=3).
 }\label{fig:layout_parametrisation}
\end{figure}
The degree of the spline $k$ is a design parameter (here set, but could also be optimised) that controls the smoothness of the cable path. For example, a linear spline ($k=1$) produces a piecewise linear path with non-differentiable corners at knots, while a cubic spline ($k=3$) produces a smooth path as shown in Figure \ref{fig:layout_parametrisation}. The choice of spline degree therefore depends on the specific application, and further practical considerations may dictate the desired level of smoothness such as ensuring a level of ease and speed when laying out the cable according to the final design specification.

When deploying the optimised design in the field, the exact replication of channel locations is of course not feasible. This is, however, unlikely to cause a problem, since small deviations from the optimised layout will typically not lead to a significant performance degradation.

Another important aspect is the choice of channel spacing when designing a survey, which ideally should be the same as the desired channel spacing in the field. However, many design criteria have a computational cost that is quadratic or worse in the number of channels (e.g., in the number of raypaths in ambient noise travel time tomography $N_\text{rays} = N_\text{cha} (N_\text{cha}-1)/2$, eigenvalue decomposition for dense matrices \citep{Golub2013-MatrixComputations}, evaluation of Gaussian data likelihood with full covariance matrices \citep{Murphy2022-ProbabilisticIntroduction}) or are just intrinsically computationally expensive (e.g., Nested-Monte Carlo for Bayesian experimental design). There is therefore an incentive to use a larger channel spacing in the design process to reduce the computational cost of the optimisation process. In some cases, the increase can be justified physically (e.g., travel times of seismic events recorded at channels close to each other provide little additional azimuthal and radial coverage, and often have correlated errors due to similar source-receiver raypaths). In other cases, it is purely a computational convenience to make the design process efficient, in which case care must be taken to ensure that the channel spacing is still sufficiently small for the chosen application. There is usually little incentive to choose an optimisation channel spacing that is smaller than the gauge length, since successive channels will not then provide independent data \citep{Kennett2024-GuideDAS}.

\subsection{Design Criteria} \label{sec:design-criteria}
Many different design goals might be considered when designing a DAS cable layout. In this study, we present computationally efficient design criteria, mostly from previous literature, that can be used to quantify common design goals. All these criteria share the characteristic that they use derived properties, such as phase picks or ambient noise correlations, instead of the full recordings of the seismic wavefield, simply because the data space (number of data points) required to describe seismic wavefields is enormous \citep[e.g.,][]{Krampe2021-OptimizedCosts,Maurer2017-OptimizedImaging}, which makes it extremely difficult to find optimal designs that are robust to nonlinearity in the physics relating wavefields to the parameters of interest \citep{Guest2010-OptimalReservoirs}. Nevertheless, the derived data space properties that we consider here cover a wide range of practical applications, while remaining sufficiently computationally efficient to allow for the global optimisation of the cable layout.

\subsubsection{Preliminaries on Signal Decay and Sensitivity}

Accounting for the signal strength of both seismic waves and laser signals along the cable is crucial when designing a DAS survey. To avoid the need for precise knowledge of noise levels and source magnitudes, we adopt a relative approach. We define a reference distance $R_\text{ref}$ at which a signal arriving at a perfectly aligned cable (one oriented along the highest angular sensitivity) is assumed to have a signal-to-noise ratio (SNR) of one. Given this reference distance, three main factors determine the SNR: (1) the distance between the cable and a seismic source, (2) the distance along the cable from the interrogator, and (3) the orientation of the incident seismic wavefield relative to the cable.

\begin{figure}
 \centering
 \includegraphics[width=0.6\textwidth]{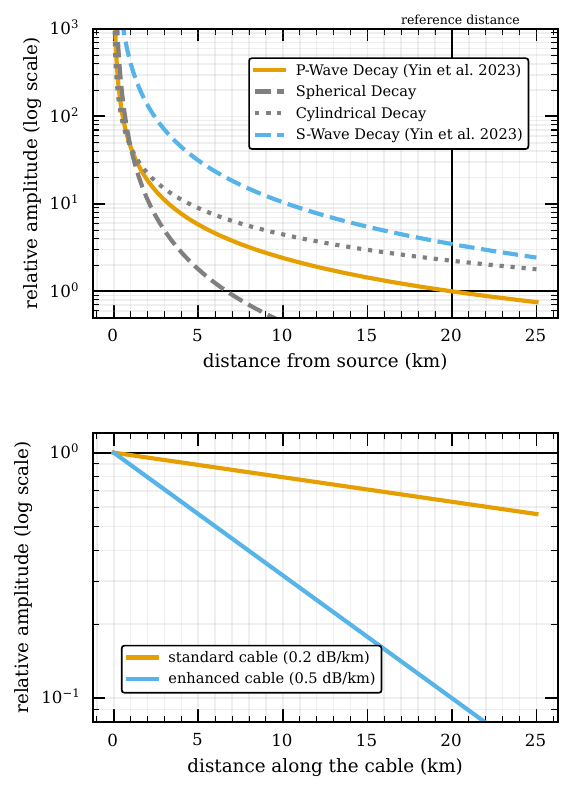}
 \caption{
  Top panel: Reduction in amplitude of seismic waves as a function of distance from source. Cylindrical decay of surface waves and spherical decay of body waves are shown alongside P- and S-wave decay derived from the local magnitude scale of \citet{Yin2023-EarthquakeRelation}.
  Bottom panel: Attenuation of laser signal as a function of distance along the cable.
 }
 \label{fig:attenuation_decay}
\end{figure}
Without prior knowledge of seismic energy attenuation due to scattering or conversion to heat, we can assume that geometric spreading dominates the attenuation of energy \citep{Baird2025-OceanNetworks}. The seismic wavefield amplitude as a function of distance $r$ from the source is then proportional to $1/r$ for spherically-spreading body waves and $1/\sqrt{r}$ for cylindrically-spreading surface waves (Figure \ref{fig:attenuation_decay}). While this provides a reasonable approximation, local magnitude scales enable more accurate modelling of the attenuation of seismic body wave phases. Although most magnitude scales are developed for traditional seismic sensors \citep{Hutton1987-TheMLscaleCalifornia} and are not directly applicable to DAS, recent studies present empirical relationships between local magnitude and source distance for DAS \citep{Barbour2017-DynamicCharacterization, Barbour2021-EarthquakeStrain,Yin2023-EarthquakeRelation,Lior2021-DetectionSensing, Lior2023-MagnitudeWarning,Baba2024-SeismicJapan}. For P- and S-waves, we use the scale of \citet{Yin2023-EarthquakeRelation}
\begin{align}
 \log_{10}(A^\text{P}_i) & = 0.437 M - 1.269 \log_{10}(R) + K^\text{P}_i \\
 \log_{10}(A^\text{S}_i) & = 0.690 M - 1.588 \log_{10}(R) + K^\text{S}_i
\end{align}
where $A^\text{P}_i$ and $A^\text{S}_i$ are the peak amplitude of the DAS strain rate for channel $i$ in microstrain/s ($10^{-6}~\text{s}^{-1}$) for P- and S-waves respectively, $M$ is the source magnitude, $R$ is the hypocentral distance in kilometres, and $K^\text{P}_i$ and $K^\text{S}_i$ are channel-specific factors that account for all local effects such as cable construction, installation, instrumental coupling, and the variety of regional geology (see Figure \ref{fig:attenuation_decay}). The scale has been proven to be transferable between different settings after only minor calibration. If we use both P- and S-waves together, we need to ensure that they are scaled consistently. We do this by calculating a reference magnitude $M_\text{ref}$ such that a P-wave at distance $R_\text{ref}$ with perfect alignment to the cable has an SNR of one, and then use this magnitude to calculate the S-wave amplitude at the same distance (this is done in Figure \ref{fig:attenuation_decay}, which results in the higher S-wave amplitude).

The laser signal propagating through the cable also experiences attenuation. For typical DAS cables, this attenuation is approximately $0.2~\text{dB/km}$ \citep{Westbrook2020-EnhancedBackscattering}, resulting in a signal decay of $0.4~\text{dB/km}$, since the reflected signal must travel the distance twice (see Figure \ref{fig:attenuation_decay}). We assume a linear relationship between laser signal strength and SNR, consistent with the Figures of \citet{Miao2025-AssessingJapan}. Enhanced scattering fibres can exhibit significantly higher attenuation values of $0.4-0.6~ \text{dB/km}$ \citep{Masoudi2021-152Fiber, Westbrook2020-EnhancedBackscattering}.

\begin{figure*}
 \centering
 \includegraphics[width=1.0\textwidth]{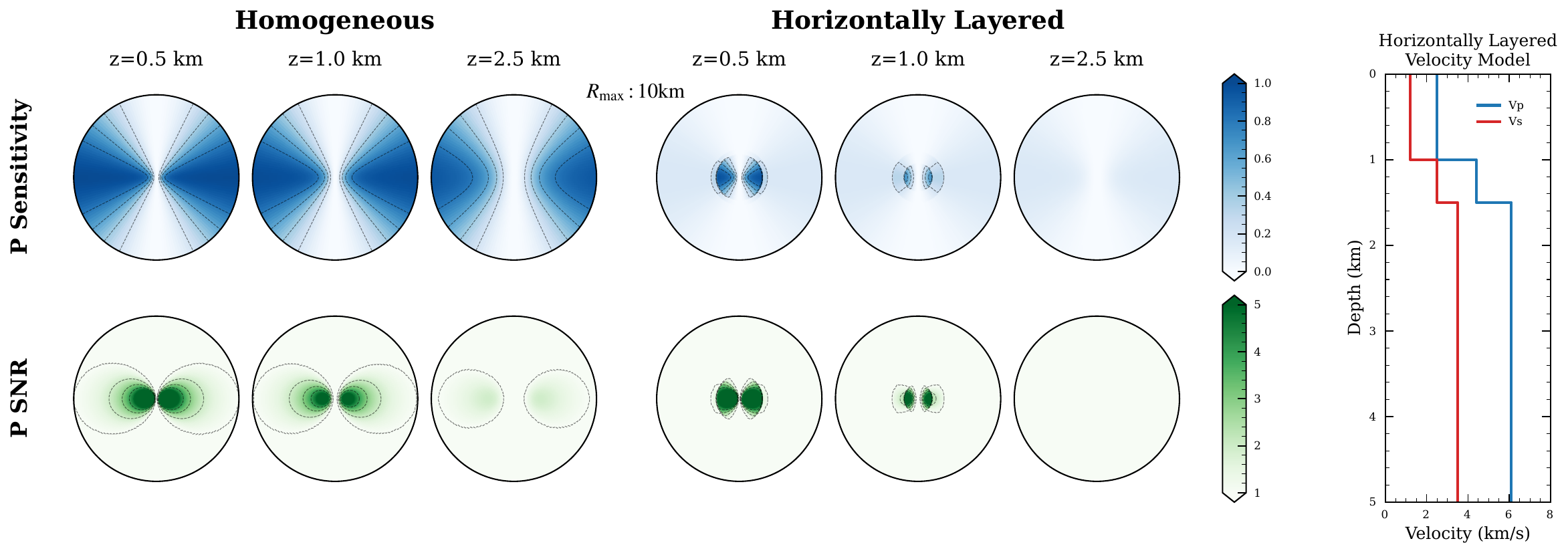}
 \caption{
  Sensitivity (top) and SNR (bottom) for P-waves as a function of azimuth and distance to the source, for sources at different depths z in both homogeneous (left) and horizontally layered (right) velocity models. The SNR calculations assume a reference distance of $R_\text{ref} = 20 \text{km}$ using the local magnitude scale of \citet{Yin2023-EarthquakeRelation}. The radius of each circle is set to $R_\text{ref}$. The channel receiver is positioned at the centre and oriented along the east-west direction for a horizontal cable. To the right, the layered velocity model used is shown.
 }\label{fig:sensitivity_snr_Pwaves}
\end{figure*}

\begin{figure*}
 \centering
 \includegraphics[width=1.0\textwidth]{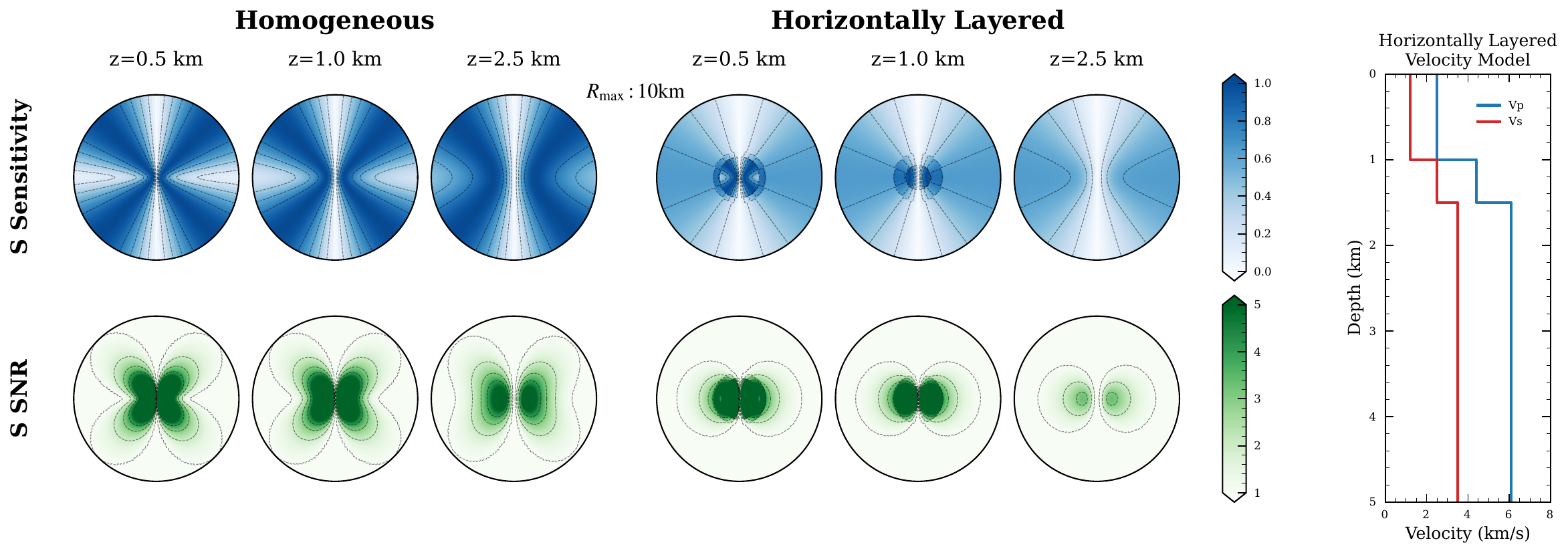}
 \caption{
  Sensitivity (top) and SNR (bottom) for S-waves as a function of azimuth and distance to the source, for sources at different depths z in both homogeneous (left) and horizontally layered (right) velocity models. The SNR calculations assume a reference distance of $R_\text{ref} = 20 \text{km}$ using the local magnitude scale of \citet{Yin2023-EarthquakeRelation}. The radius of each circle is set to $R_\text{ref}$. The channel receiver is positioned at the centre and oriented along the east-west direction for a horizontal cable. To the right, the layered velocity model used is shown.
 }\label{fig:sensitivity_snr_Swaves}
\end{figure*}

The primary distinction between single-component strain measurements along a DAS cable and traditional seismic sensor measurements lies in how cable orientation affects wavefield sensitivity, as well as in the measurement of strain rate rather than particle velocity. In our analysis of this angular sensitivity, we exclude wave amplitude factors and frequency or phase dependencies, as these characteristics are typically unknown for prospective arrivals prior to detection \citep{Hudson2025-TowardsNetworks}.

The strain rate sensitivity $\xi$ quantifies how the cable orientation affects the measured amplitude relative to the true seismic wave amplitude. It is defined as
\begin{equation}
A_\text{meas} = \xi \cdot A_\text{wave}
\end{equation}
where $A_\text{meas}$ is the amplitude measured by the DAS cable and $A_\text{wave}$ is the amplitude of the incident seismic wave. The sensitivity $\xi$ ranges from 0 (no sensitivity) to 1 (maximum sensitivity) depending on the wave type and cable orientation. Constant multipliers in the following equations are removed since we use relative SNRs. For P, SH and SV waves, the sensitivity is given by \citep{Martin2017-SensitivityArrays,Hudson2025-TowardsNetworks}
\begin{align}
 \xi_\text{P}  & = \left\lvert \cos^2 \left(\theta \right) \sin^2 \left( \varphi \right) \right\rvert \\
 \xi_\text{SH} & = \left\lvert  \cos^2 \left(\theta \right) \sin \left(2 \varphi \right) \right\rvert \\
 \xi_\text{SV} & = \left\lvert  \sin \left(2\theta \right) \cos \left(\varphi \right) \right\rvert
\end{align}
where $\theta$ represents the azimuth of the channel relative to the wave propagation direction on a plane in which the cable lies, and $\varphi$ is the incidence angle, which is the angle of the plane wave relative to the plane-perpendicular angle (e.g., angle from vertical-down for a channel on a horizontal section of cable). Following \citet{Hudson2025-TowardsNetworks}, we define the S-wave sensitivity as
\begin{equation}
 \xi_\text{S} = \max \left( \sqrt{\frac{1}{2} \left( \xi_\text{SH}^2 + \xi_\text{SV}^2 \right)}, \xi_\text{SV}, \xi_\text{SH} \right)
\end{equation}
which allows us to combine the two S-wave polarisation modes.

This anisotropic sensitivity creates complex patterns in the SNR as a function of azimuth and distance to the source. Figure \ref{fig:sensitivity_snr_Pwaves} illustrates the directional sensitivity and SNR for P-waves across varying azimuths, distances and source depths for a horizontal cable. If the cable was dipping significantly, the sensitivity would become asymmetric. The SNR calculations assume a reference distance of $R_\text{ref} = 10 \text{km}$ and use the local magnitude scale of \citet{Yin2023-EarthquakeRelation}. The absence of sensitivity to broadside arrivals is evident, with the insensitive area below the channel widening as depth increases. The velocity model significantly affects body wave phase sensitivity, as velocity that increases with depth generally results in smaller inclination angles and consequently reduced sensitivity.

Figure \ref{fig:sensitivity_snr_Swaves} shows the equivalent analysis for S-waves. Rather than a source with equal source magnitude, the same reference distance as for P-waves is used to allow direct comparison. S-wave sensitivity is generally higher but exhibits a more complex pattern (dominated by the $\sin(2\theta)$ term for small incidence angles, and a combination of $\sin(2\theta)$ and $\cos^2(\theta)$ for larger ones). The layered velocity model causes dramatic changes in sensitivity patterns whenever the first arrival changes to critically refracted waves which travel along a deeper interface. For a source at a depth of 0.5~km, this even creates a small region of zero sensitivity within the main lobe of the S-wave sensitivity pattern (see Figure \ref{fig:sensitivity_snr_Swaves} panels for a source at 5 km depth). In practical applications, sensitivity and SNR patterns will deviate from the idealised patterns shown here and may be smoother due to heterogeneous velocity structures and phase conversions. Both factors could be accounted for by using more complex velocity models or smoothing the sensitivity and SNR patterns.

\begin{figure*}
 \centering
 \includegraphics[width=1.0\textwidth]{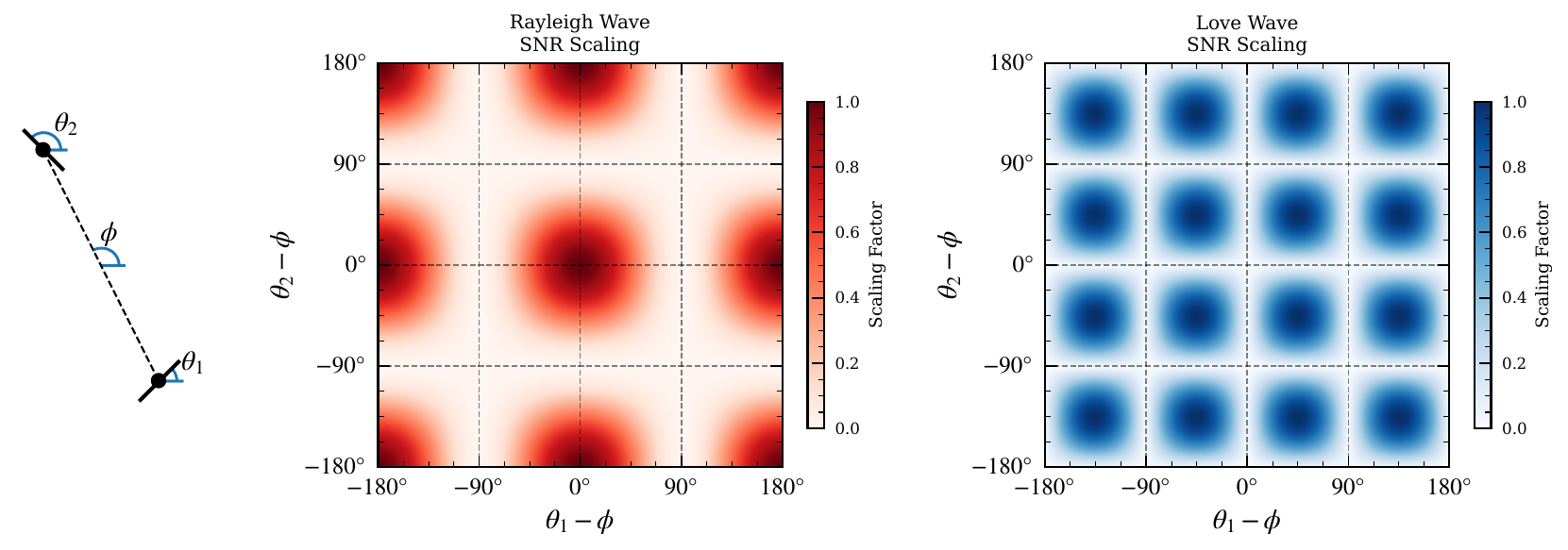}
 \caption{
  Sensitivity for Rayleigh (center) and Love (right) waves as a function of the azimuths of the two channels relative to the wave propagation direction, which is set to have azimuth zero for the purposes of display. The sketch on the left illustrates the azimuths of the two channels $\theta_1$ and $\theta_2$ and the direction of the line connecting the two channels $\phi$. Adapted from \citet{Fang2023-DirectionalCalifornia}.
 }\label{fig:rayleigh_love_snr_scaling}
\end{figure*}

Similar to body waves, surface waves exhibit anisotropic sensitivity described by \citep{Martin2021-IntroductionMeasurements, Martin2017-SensitivityArrays}
\begin{align}
 \xi_\text{rayleigh} & =  \left\lvert \cos^2 \left( \theta \right) \right\rvert \label{eq:rayleigh_sensitivity} \\
 \xi_\text{love}     & =  \left\lvert \sin \left( 2 \theta \right) \right\rvert \label{eq:love_sensitivity}
\end{align}
where $\theta$ is the azimuth of the channel relative to the wave propagation direction. Rayleigh wave sensitivity is therefore equivalent to P-wave sensitivity for a wave that propagates in the same plane as the cable, while Love wave sensitivity corresponds to SV-wave sensitivity for a wave that propagates in the same plane as the cable.

An interesting case arises for ambient noise analysis that involves cross-correlation between pairs of channels in the horizontal plane with azimuths $\theta_1$ and $\theta_2$, where $\phi$ represents the direction of the line that connects these channels. The ambient noise cross-correlation can be used to estimate the average group velocity between channels \citep{Curtis2006-SeismicSignal} to perform ambient noise surface wave tomography \citep{Shapiro2005-High-resolutionNoise, Fang2023-DirectionalCalifornia}, which requires no active source and can therefore be used to estimate the subsurface velocity structure passively. The approach is based on the assumption that the cross-correlation of ambient noise between two channels is proportional to the Green's function between these channels, at least for the wave phases or properties of interest \citep{Snieder2004-ExtractingPhase, Shapiro2005-High-resolutionNoise}. 

The signal amplitude in the cross-correlation can be described by \citep{Fang2023-DirectionalCalifornia, Martin2021-IntroductionMeasurements}
\begin{align}
 \xi_\text{rayleigh-cc} & = \left\lvert \cos^2 \left( \theta_1 - \phi \right) \cos^2 \left( \theta_2 - \phi \right) \right\rvert \\
 \xi_\text{love-cc}     & = \left\lvert \sin 2 \left( \theta_1 - \phi \right) \sin 2 \left( \theta_2 - \phi \right) \right\rvert
\end{align}
as illustrated in Figure \ref{fig:rayleigh_love_snr_scaling}. For surface wave tomography that uses earthquake data, Equations \ref{eq:rayleigh_sensitivity} and \ref{eq:love_sensitivity} could be applied directly.

Several approximations were made in deriving the azimuthal dependencies: effects of wavelength, gauge length, and local velocity structure were neglected. While beyond this study's scope, analytical expressions exist for some more complex cases \citep{Martin2021-IntroductionMeasurements, Martin2017-SensitivityArrays}, allowing straightforward incorporation of this information in cases for which it is available. The azimuthal dependencies of sensitivity and SNR presented here therefore represent first-order approximations that should be used with caution, but which nevertheless provide valuable insights for DAS cable layout design, as many of the incorporated effects are observed in field data \citep[e.g.,][]{Feng2023-EFWIProperties,Baird2025-OceanNetworks, Mateeva2014-DistributedVSP,Yin2023-EarthquakeRelation}.

\subsubsection{Linearised Travel Time Tomography}
Linearised travel time tomography is a tool for imaging subsurface velocity structure using travel time measurements between pairs of receivers or between sources and receivers. It relies on the assumption that ray paths can be approximated as fixed, allowing the inverse problem to be formulated as a linearised system of equations:
\begin{equation}
 \mathbf{d} = \mathbf{A} \mathbf{m} + \boldsymbol{\epsilon}
\end{equation}
where $\mathbf{d}$ is the vector of observed data (inter-channel time lags, or equivalently travel times), $\mathbf{m}$ is the model vector (group or phase velocity at a set of locations), $\mathbf{A} = {\partial d_i}/{\partial m_j}$ is the sensitivity matrix of partial derivatives of each datum $d_i$ with respect to each model parameter $m_j$, and $\boldsymbol{\epsilon}$ is the noise vector. If the velocity model is parameterised on a grid of non-overlapping cells, $\mathbf{A}$ can be constructed from the ray paths between channels as $A_{ij} = l_{ij}$, where $l_{ij}$ is the length of ray path $i$ in cell $j$. Here, each ray defines the inter-channel path followed by the wave packet for which travel time data $d_i$ were measured. While in our examples we follow the common approach of designing surveys using a homogeneous velocity model to calculate $\mathbf{A}$ \citep{Fang2023-DirectionalCalifornia}, in general the best prior estimate of the velocity model in any particular survey area should be adopted.

\begin{figure*}
 \centering
 \includegraphics[width=1.0\textwidth]{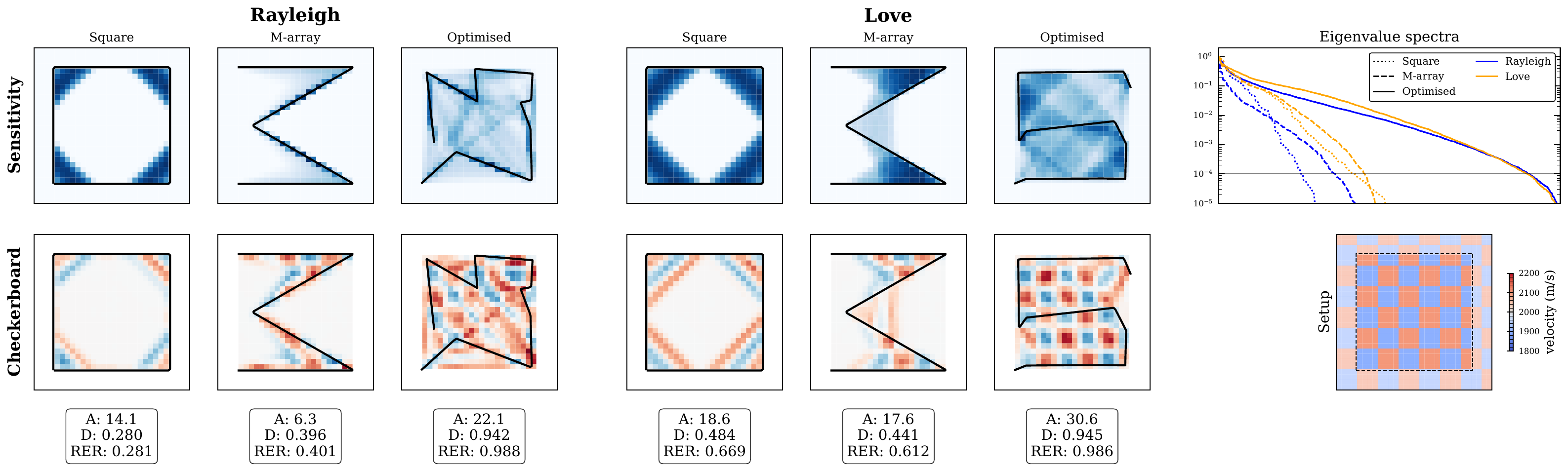}
 \caption{
  Schematic illustration of how different cable layouts (all of the same length) affect various measures of information about the model provided by recorded Rayleigh (left) and Love (right) wave data. The top row shows the average over all data of the sensitivity to each velocity grid cell for a checkerboard model; the bottom row shows the results of inverting a checkerboard model using the sensitivity matrix $\mathbf{A}$ for each layout. The side length of the plots is 3000~m and the reference distance used for the attenuation of seismic energy is 7500~m for Rayleigh waves and 2500~m for Love waves, which are chosen in a way that the different layouts produce a range of eigenvalue spectra. The normalised eigenvalue spectrum is shown top right, and the true velocity model is shown bottom right with the square array overlain for reference. Normalised A-optimality $\Sigma_A^*$, D-optimality $\Sigma_D^*$ ($\lambda_\text{thr} = 10^{-10}$), and relative eigenvalue range (RER) $\Sigma_\text{RER}$ ($\lambda_\text{thr} = 10^{-5}$) values are provided below each layout labeled A, D and RER respectively. The optimised layouts are D-optimal.
 }\label{fig:tomo_schematic}
\end{figure*}

The matrix $\mathbf{A}$ determines the sensitivity of the data to the underlying velocity model, and its eigenvalue spectrum can be used to quantify the inversion stability and information transfer from data to parameter estimates \citep{Menke2018-GeophysicalTheory}. Since measurement errors propagate into the solution $\mathbf{m}$ parallel to each eigenvector of $\mathbf{L} = \mathbf{A}^T \mathbf{A}$ with an amplification of $1/\lambda_i$, where $\lambda_i$ is the $i$-th eigenvalue of $\mathbf{L}$, optimal experimental design for linearised methods aims to maximise eigenvalues of $\mathbf{L}$. Figure \ref{fig:tomo_schematic} illustrates how different cable layouts affect the eigenvalue spectrum for both Rayleigh and Love waves. For each design, we show the average sensitivity to each velocity grid cell across all data, and demonstrate the effect on inversion results with a checkerboard model. Data for the checkerboard inversion were generated with a seismic ray-tracer \citep{Giroux2021-TtcrpyRaytracing}, which accurately models the ray paths in the heterogeneous checkerboard model, and were inverted using the sensitivity matrix based on a homogeneous prior model. Layouts with high eigenvalue spectra exhibit both better sensitivity coverage and improved resolution of the checkerboard model.

Several approaches exist for quantifying eigenvalue positivity in optimal experimental design \citep{Atkinson1992-OptimumDesigns, Curtis2004-TheoryProblems-01}. A-optimality \citep{Curtis1997-ReconditioningParameterization,Krampe2021-OptimizedCosts}, D-optimality \citep{Box1959-DesignSituations,Curtis1999-OptimalSurveys,Curtis1999-OptimalExamples,Kijko1977-AlgorithmNetwork--I,Kijko1977-AlgorithmStations,Rabinowitz1990-OptimalApproach,Steinberg1995-ConfiguringSources}, and the relative eigenvalue range (RER) \citep{Maurer2009-FrequencyExperiments} are the most commonly used criteria in geophysical applications.

In the examples in this study, we use the D-optimality criterion $\Sigma_D$, which is defined as
\begin{align}
 \Sigma_D = \left| \det(\mathbf{L}) \right| = \prod_{i=1}^N \lambda_i = \sum_{i=1}^N \log(\lambda_i) \label{eq:D-optimality}
\end{align}
and relates directly to the volume of the posterior covariance matrix \citep{Box1959-DesignSituations}. To allow the criterion to be applied to underdetermined problems, a threshold $\lambda_\text{thr}$ is often applied \citep{Curtis1999-OptimalExamples}, below which eigenvalues are set to a penalty term $\lambda_\text{pen}$. The penalty term should be a sufficiently large negative number to ensure that almost all possible logarithm of eigenvalues exceed $\lambda_\text{thr}$. The D-optimality criterion in Equation \ref{eq:D-optimality} measures the volume under the logarithmic eigenvalue spectrum, but for ease of interpretation, in mixed or under-determined problems, we define an adapted normalised D-optimality criterion $\Sigma_D^*$ as
\begin{align}
 \Sigma_D^* = 1 - \frac{\sum_{i: \lambda_i \geq \lambda_\text{thr}} \log\left(\frac{\lambda_i}{\lambda_{\max}}\right) + N_{\text{below}} \cdot \lambda_\text{pen}}{N \cdot \lambda_\text{pen}}
\end{align}
where $\lambda_{\max}$ is the maximum eigenvalue, and $N_{\text{below}}$ is the number of eigenvalues below the threshold. This measure is bounded between 0 and 1, with 1 indicating a perfect design where all eigenvalues are equal to $\lambda_{\max}$, and 0 indicating a design where all eigenvalues are below the threshold $\lambda_\text{thr}$. Importantly, this represents only a different formulation of the (normalised) D-optimality criterion and does not change the optimal design.

The threshold and penalty term are necessary due to the ill-conditioning of the sensitivity matrix $\mathbf{L}$ \citep{Curtis1999-OptimalExamples}. If the sensitivity matrix is regularised by introducing prior knowledge in the form of an a priori model covariance matrix $\mathbf{C}_m$ and a data covariance matrix $\mathbf{C}_d$ \citep{Maurer2010-RecentDesign}, the original D-optimality criterion can be used without the need for a penalty term. While including such information lies beyond the scope of this design methodology study (since it involves case-dependent prior information) it is important to note that such prior information can make the optimisation process more robust and allows it to focus better on the least constrained parts of the model according to the prior information.

The A-optimality criterion is computationally more efficient as it does not require the full eigenvalue spectrum to be calculated: the sum of eigenvalues is equal to the trace of $\mathbf{L}$, and the largest eigenvalue (used for normalisation) can be estimated using the power method \citep{Curtis1997-ReconditioningParameterization,Curtis1999-OptimalExamples,Curtis1999-OptimalSurveys}. This efficiency comes at the cost of often leading to designs with a small number of large eigenvalues at the expense of many small eigenvalues. For an example of this, see the Rayleigh wave results in Figure \ref{fig:tomo_schematic}, where the square design performs reasonably well according to the A-optimality criterion, but the checkerboard model is poorly resolved, which is in line with the findings of \citet{Krampe2021-OptimizedCosts} in the context of full waveform inversion. In this study, the scenarios considered are small enough to calculate the full eigenvalue spectrum and therefore employ the adapted D-optimality criterion $\Sigma_D^*$ as our design criterion. For a detailed discussion of linearised optimal experimental design for seismic tomography, we refer to \citet{Curtis2004-TheoryProblems-01}.

We include attenuation effects and anisotropic sensitivity by first calculating all ray paths between all pairs of channels, and then removing those for which the SNR is below a threshold value. The SNR depends on the inter-channel distance and alignment and is set to one for channels with perfect alignment at a distance of $R_\text{ref}$.

In recent decades, several improvements to linearised experimental design have been proposed. These advances account for prior knowledge in a Bayesian, yet still linearised, manner \citep[e.g.,][]{Englezou2022-ApproximateModels,Long2015-BayesianInversion,Alexanderian2021-OptimalReview}, and enhance computational efficiency for large-scale problems \citep[e.g.,][]{Mercier2025-DesigningApproach,Wu2023-Large-scaleNetwork, Wu2023-Offline-onlinePlacement,Wu2023-ScalableDesign}, or aim to increase robustness against model uncertainty \citep{Alexanderian2022-OptimalUncertainty} and worst-case scenarios \citep{Chowdhary2024-RobustProblems}. Since this study focuses on DAS cable layout optimisation and the computational cost is relatively low for the scenarios considered here, we do not apply these methods here, but they can be directly integrated into our optimisation process when the required prior information is available.

\subsubsection{Probabilistic Source Location}
Linearised experimental design, as introduced in the previous section, efficiently optimises experimental designs but is inherently limited by its reliance on linearisation and cannot account for the full non-linearity in the physics of the problem \citep{Curtis2004-TheoryProblems,Strutz2024-VariationalReservoir}. As an example of fully non-linear Bayesian experimental design, we consider the optimisation of a DAS layout for probabilistic source location. In this approach, we aim to estimate the posterior probability distribution of the source location $\mathbf{m}$ given the observed data $\mathbf{d}$ and design $\boldsymbol{\xi}$
\begin{equation}
 p(\mathbf{m} \, | \, \mathbf{d}, \boldsymbol{\xi}) = \frac{p(\mathbf{d} \, | \, \mathbf{m}, \boldsymbol{\xi}) \, p(\mathbf{m})}{p(\mathbf{d} \, | \, \boldsymbol{\xi})}
\end{equation}
where $p(\mathbf{d} \, | \, \mathbf{m}, \boldsymbol{\xi})$ is the data likelihood given the source location and design, and $p(\mathbf{m})$ is the prior distribution of the source location. The prior distribution describes our belief about where seismicity is most likely to occur in the subsurface before any data observation from the proposed survey, while the likelihood quantifies the probability of observing the recorded data given the model parameters and experimental design.

The data likelihood function should take the form of expected observational uncertainties, so we define it as a multivariate Gaussian distribution
\begin{equation}
 p(\mathbf{d} \, | \, \mathbf{m}, \boldsymbol{\xi}) = \mathcal{N}( \mathrm{F} \left(\mathbf{m}, \boldsymbol{\xi} \right), \boldsymbol{\Sigma}_d)
\end{equation}
where $\mathrm{F}(\mathbf{m}, \boldsymbol{\xi})$ is the forward function that predicts arrival times for a given source location and design, and $\boldsymbol{\Sigma}_d$ is the expected data covariance matrix. The forward function typically involves ray tracing to calculate expected arrival times of seismic phases at channels based on the source location and velocity model.

The data covariance matrix incorporates the effects of signal decay and sensitivity discussed earlier, and can be modelled as:
\begin{align}
 \boldsymbol{\Sigma} & = \boldsymbol{\Sigma}_\text{pick} + \boldsymbol{\Sigma}_\text{vel} + \boldsymbol{\Sigma}_\text{obs}
\end{align}
where $\boldsymbol{\Sigma}_\text{pick}$ represents the uncertainty in phase picking, $\boldsymbol{\Sigma}_\text{vel}$ accounts for uncertainty in the velocity model, and $\boldsymbol{\Sigma}_\text{obs}$ quantifies general observation noise such as channel location uncertainty or other measurement errors \citep{Strutz2025-RolesDioxide}. The first two terms are functions of the source location $\mathbf{m}$ and the design $\boldsymbol{\xi}$, while the third term is typically a constant value.

We convert the SNR into picking uncertainty $\boldsymbol{\Sigma}_\text{pick}$ via an information-theoretic approach \citep{Fuggi2024-AssessmentNetworks, Aki1976-SignalMeasurements}:
\begin{equation}
 \left(\Sigma_\text{pick} \right)_{ii}= \left[
  \log_2 \left( 1 + \frac{20 \log_{10} \text{SNR} \left(\mathbf{m}, \boldsymbol{\xi}\right)_i}{K} \right) \, 2 \, f_\text{max}
  \right]^{-2}  \label{eq:Shannon_Hartley}
\end{equation}
where $\text{SNR}(\mathbf{m}, \boldsymbol{\xi}) = A_\text{signal}(\mathbf{m}, \boldsymbol{\xi}) \,  /  \, A_\text{noise}(\boldsymbol{\xi})$ is the signal-to-noise ratio of the signal amplitude $A_\text{signal}(\mathbf{m}, \boldsymbol{\xi})$ and the noise amplitude $A_\text{noise}(\boldsymbol{\xi})$, and $f_\text{max}$ is the maximum frequency of the recorded signal. The absolute value of $A_\text{noise}(\boldsymbol{\xi})$ need not be known since the reference distance $R_\text{ref}$ normalises the SNR. Relative changes in $A_\text{noise}(\boldsymbol{\xi})$, for example due to local conditions or noise sources, can be included by multiplying the SNR of a given channel by a factor that accounts for the change in noise level. The empirical factor $K=10$ in Equation \ref{eq:Shannon_Hartley} is chosen to align with empirical estimates of picking uncertainty from \citet{Zelt1994-ModelingProvince} and \citet{PabloCanales2000-SeismicRidges}, and differs from the value used by \citet{Fuggi2024-AssessmentNetworks} and \citet{Aki1976-SignalMeasurements}. The exact values of both $K$ and the maximum frequency $f_\text{max}$ are not critical for design optimisation, as they scale the picking uncertainty uniformly across all channels while preserving the key characteristic of rapidly increasing uncertainty at low SNR values.

For the velocity model uncertainty, we assume that the variance of the travel time errors is proportional to the travel time itself:
\begin{align}
 \left( \Sigma_\text{vel} \right)_{ii} = d_{i} \cdot \varsigma_\text{vel}^2 \label{eq:velocity_model_uncertainty}
\end{align}
where $d_i$ is the travel time of the $i$-th phase arrival and $\varsigma_\text{vel}$ is a scaling factor that determines the magnitude of uncertainty we assume in the seismic velocity model. This assumption is based on a Gaussian random walk model, where the variance of the distance from a reference point (in our case, the mean arrival time) is proportional to the time (here arrival time). Intuitively, this means that the longer the travel time, the more uncertainty it accumulates along its path.

While incorporating velocity model knowledge into the forward model is straightforward (albeit potentially computationally expensive), quantifying arrival time uncertainty is more complex. The approach presented here is one of many possible methods, with the optimal choice depending on the specific application and available prior information. Alternative approaches in the literature include using a linear relationship between arrival time and its standard deviation \citep{Fuggi2024-AssessmentNetworks,Tarantola1982-InverseInformation}, polynomial fits that describe uncertainty as a function of source-receiver distance and depth \citep{Callahan2025-AnalysisDesign}, or (heuristic) estimates based on the attenuation of seismic energy \citep{Curtis2004-DeterministicSurveys}.

In equation \eqref{eq:velocity_model_uncertainty}, we have assumed that the measurement uncertainties at different receivers are independent of each other. This is not the case if the seismic wave paths to nearby receivers travel similar routes through the subsurface, therefore accumulating similar deviations. While exactly quantifying this dependence is difficult, we use a heuristic approach \citep{Tarantola1982-InverseInformation, Callahan2025-AnalysisDesign} and base the correlation of the uncertainties on the inter-receiver distances $l_{ij}$ defined as:
\begin{equation}
 C_{ij} = \exp{\left( -\frac{1}{2}\frac{l_{ij}^2}{l_\text{cor}^2} \right)}
\end{equation}
where $l_\text{cor}$ is the correlation length. If correlations are expected, the covariance matrix becomes:
\begin{equation}
 \hat{\boldsymbol{\Sigma}}_\text{vel} = \boldsymbol{\Sigma}_\text{vel}(\mathbf{m}, \boldsymbol{\xi}) \cdot \mathbf{C}
\end{equation}

As a design criterion for probabilistic source location, we use the expected information gain (EIG) \citep{Lindley1956-MeasureExperiment}:
\begin{align}
 \operatorname{EIG} & = \mathbb{E}_{p(\mathbf{d} | \boldsymbol{\xi})} \Bigl\{ \operatorname{I}[p(\mathbf{m} \, | \, \mathbf{d},\boldsymbol{\xi})] - \operatorname{I}[p({\mathbf{m}})] \Bigr\}    \\
                    & =  \mathbb{E}_{p(\mathbf{m})} \Bigl\{ \operatorname{I}[p(\mathbf{d} \, | \, \mathbf{m},\boldsymbol{\xi})] - \operatorname{I}[p({\mathbf{d}}| \, \boldsymbol{\xi})] \Bigr\}
 \label{eqn_03:EIG}
\end{align}
where $\operatorname{I}[p]$ denotes the Shannon information \citep{Shannon1948-MathematicalCommunication,Cover2006-ElementsTheory} of distribution $p$. The EIG quantifies the expected reduction in uncertainty about the source location after we observe the data, compared to its prior uncertainty. We calculate the EIG via the robust nested Monte Carlo (NMC) method \citep{Ryan2003-EstimatingModel,Huan2013-Simulation-basedSystems} to avoid assumptions about the posterior distribution. For detailed treatment of the EIG in geophysical applications, see \citet{Strutz2024-VariationalReservoir}.

\subsubsection{Other Design Criteria}
The design optimisation process can be adapted to any design criterion that takes a cable layout as input and returns a scalar value as output. An example of another criterion is the offset-azimuth distribution for constraining azimuthal anisotropy \citep{Fichtner2022-SimpleSensing}. The EIG used for the source location criterion could be adapted to probabilistic moment tensor estimation. This would require more prior information, such as phase amplitude models and uncertainty, and would come at a higher computational cost since more samples would be necessary to estimate the EIG.

\subsection{Design Optimisation} \label{sec:design-optimisation}

\begin{figure*}
 \centering
 \includegraphics[width=0.9\textwidth]{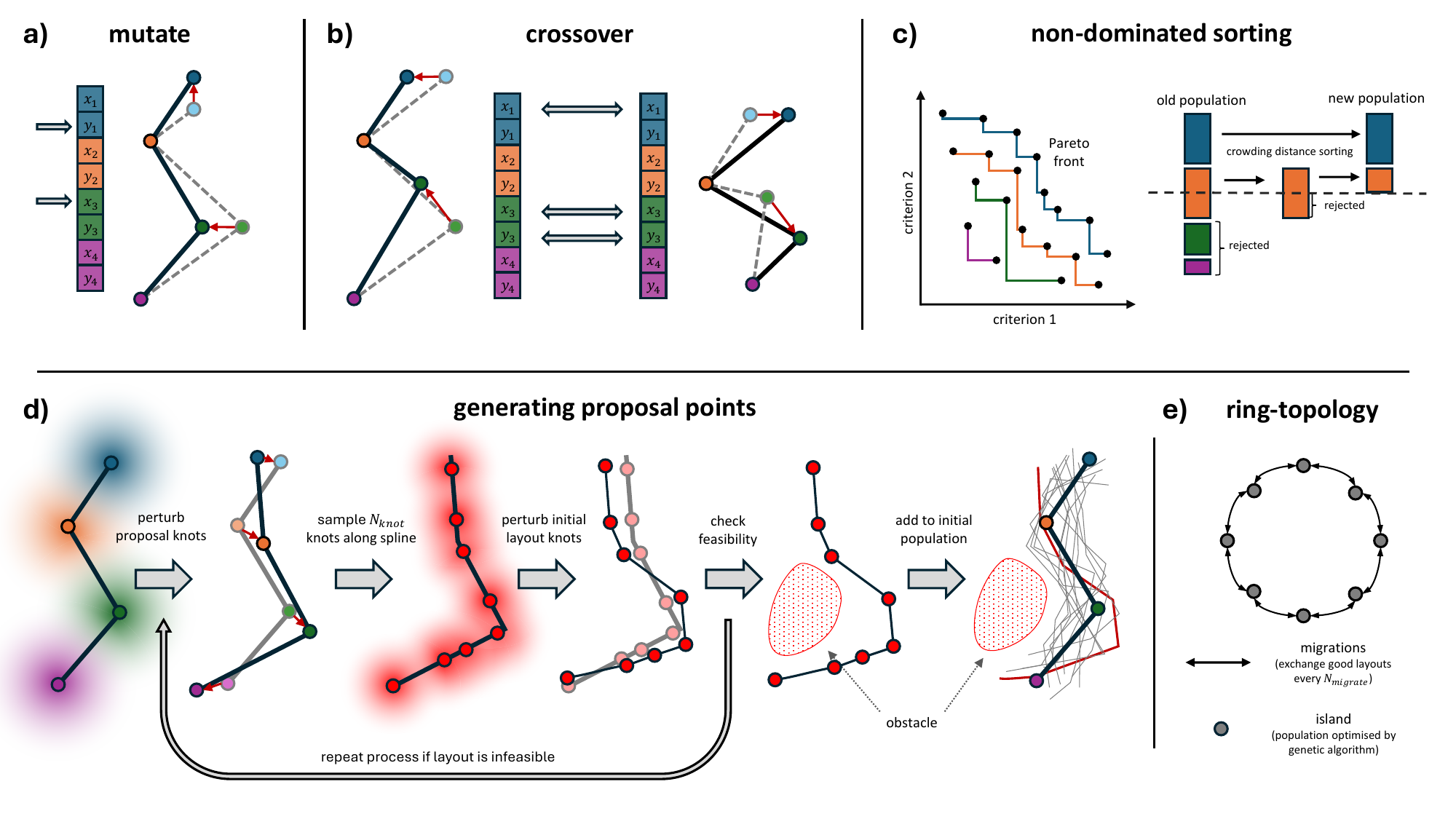}
 \caption{
  Schematic illustration of selected steps in the design optimisation process for DAS cable layouts. a) and b) respectively illustrate the mutation and crossover operations used to evolve the population of candidate layouts. c) shows a schematic example of how the non-dominated sorting algorithm works, where the layouts (black dots) are first sorted into non-dominated fronts (coloured lines) from which the best layouts are selected for the next generation. If a front exceeds the population size threshold, layouts from this front are selected based on crowding distance sorting, which ensures layouts are well distributed along the front. d) generation of the initial population from a single proposal layout. For more details, see the text. e) illustrates how layouts in a ring topology are exchanged between islands.
 }\label{fig:DAS_OED_schematic}
\end{figure*}

The optimisation of DAS cable layouts is a non-linear optimisation problem for which ideally we want to find the global optimum. Since gradient-based optimisation methods would require the design criterion to be differentiable with respect to the knots describing the cable layout, and are often prone to local optima, we choose to use derivative-free optimisation methods, specifically evolutionary algorithms \citep{Vikhar2016-evolutionaryprospects} due to their flexibility and robustness. These algorithms work by evolving a population of layouts of size $N_\mathrm{pop}$ over a number of generations. In each generation, the population is evaluated using the design criteria and the best layouts are selected and used to form the next generation. During the optimisation process, the population is mutated (partially altered) and recombined (e.g., components of pairs of designs are exchanged) to explore the design space (see Figure \ref{fig:DAS_OED_schematic}a and b for an illustration of these operations) and move towards the optimal solution. While they are not guaranteed to find the global optimum in finite time, they work well in practice for many problems and can escape local optima.

To optimise the cable layout, we optimise the knot locations $\mathbf{k}_i$, from which channel locations are interpolated using a given channel spacing and spline degree $k$ (which could also be optimised). All knots and channels must lie within a bounded area and cannot be positioned within predefined restricted areas, while the cable layouts must not exceed the maximum permissible cable length. Additionally, user-defined fixed knots between the optimisable knots can constrain the cable to pass through certain locations or follow a specific path for a portion of the cable.

An important aspect of evolutionary algorithms is the generation of the initial population. If the initial population is too similar, the optimisation process may become stuck in local optima and fail to explore the design space effectively. Conversely, if the initial population is generated completely randomly, nearly all layouts may violate design constraints such as maximum cable length and restricted areas. This is particularly problematic when design constraints are strict, making the generation of valid layouts computationally expensive.

To address this challenge, we use one or more proposal cable layouts to guide the generation of the initial population. Proposal layouts are splines with $N_\mathrm{prop}$ manually defined knots that represent reasonable starting points for the optimisation process. These layouts provide a convenient means of introducing expert knowledge and heuristics, leading to faster convergence while balancing the risk of biasing results towards the proposals.

The initial population generation process proceeds as follows (illustrated in Figure \ref{fig:DAS_OED_schematic}d). We begin with a set of proposal layouts, which need not be of the same length, and assign a normalised weight $w_\mathrm{prop}$ to each proposal. The weight controls the fraction of the initial population generated from each proposal and is typically set to $w_\mathrm{prop} = 1 / N_\mathrm{total}$, where $N_\mathrm{total}$ is the total number of proposal layouts. For each proposal layout, we create $N_\mathrm{pop} \times w_\mathrm{prop}$ initial layouts, where $N_\mathrm{pop}$ is the total population size.

To generate a single initial cable layout, we employ a two-stage perturbation process. First, we perturb the proposal knots using a Gaussian distribution with standard deviation $\epsilon_\mathrm{prop}$. Second, we randomly sample $N_\mathrm{knots}$ knot points along a spline that interpolates the perturbed proposal knots, ensuring that all layouts in the initial population have the same number of knots. These sampled knots are then further perturbed using another Gaussian distribution with standard deviation $\epsilon_\mathrm{knot}$.

This two-stage approach is important for constraint satisfaction. If perturbations were applied only to the final knot locations, the resulting cable layout could easily violate the maximum cable length constraint. The separation of proposal perturbation from knot location perturbation helps maintain feasible layouts. For layouts with many knots, perturbations can be correlated with the distance along the cable between knot points, which further helps prevent violations of the permissible cable length.

If the resulting cable layout violates any constraints, such as exceeding the maximum cable length or entering restricted areas, it is rejected and a new one is generated. This process repeats until a sufficient number of valid layouts have been created for each proposal. When design constraints are strict, only a small fraction of possible knot locations will result in permissible cable layouts. Therefore, proposal points and perturbation strengths must be chosen carefully to ensure the initial population can be generated within a computationally tractable number of iterations.

This approach balances manual input with automated generation, enabling the incorporation of expert knowledge while effectively exploring the design space. Although the method described above works well for the examples in this study, it represents one of many possible approaches for generating initial populations, ranging from completely random layouts to fully hand-crafted designs.

Once the initial process is complete, the population is evolved over several generations. Various strategies can be employed to evolve the population \citep[e.g.,][]{Gad2022-ParticleReview, Storn1997-DifferentialSpaces}, and we chose to use a genetic algorithm (GA) \citep{Holland2021-AdaptationSystems, Holland1984-GeneticAdaptation} as it performed best in our tests. The genetic algorithm evolves the population by selecting the $N_\mathrm{select}$ best layouts, applying simulated binary crossover \citep{Deb2000-ElitistNSGA-II} to combine them in random ways, introducing random mutations via polynomial mutation, and retaining the best $N_\mathrm{pop}$ layouts for the next generation. This process iterates until the maximum number of generations or another stopping criterion is reached. Early termination is often beneficial when the population shows no improvement over several generations, thus balancing solution quality with computational efficiency.

Sorting layouts by their design criterion is straightforward for single-objective optimisation. For multi-objective optimisation, we use the non-dominated sorting genetic algorithm (NSGA-II) \citep{Deb2000-ElitistNSGA-II}. This algorithm sorts the population into non-dominated (Pareto) fronts and selects the best layouts from each front, using a crowding distance to ensure diversity when necessary (see Figure \ref{fig:DAS_OED_schematic}e for an illustration). The Pareto front is the set of layouts that cannot be improved in any of the objectives without degrading at least one of the other objectives. It is constructed by identifying all layouts in the population that are not dominated by any other layouts - meaning no other solution is better in all objectives while being strictly better in at least one. The crowding distance measures how close the layouts are to each other in the design criterion space. The final Pareto front allows one to choose the best layout for each design criterion individually, or a compromise between them.

Since the design optimisation process is non-deterministic, results will vary between runs unless the random seed is fixed. To ensure robust results, we recommend running the optimisation multiple times using different seeds. We achieve this efficiently by employing the generalised island model \citep{Izzo2012-GeneralizedModel}, in which each island (a population and an evolution strategy) evolves independently, except that periodically (every $N_\text{migrate}$) it exchanges layouts with other islands in a conceptual archipelago. This exchange of information improves the convergence of the optimisation process while allowing for parallelisation with minimal communication overhead \citep{Izzo2012-GeneralizedModel}. In this study, we use a ring topology, where each island exchanges layouts only with its two neighbours (see Figure \ref{fig:DAS_OED_schematic}e for an illustration). We use the pagmo library \citep{Biscani2020-ParallelPagmo} to implement both the evolutionary algorithms and the generalised island model. If sufficient computational resources are available, islands can be set up with different evolutionary strategies to further increase the robustness of the results.

\section{Slope Instability Case Study}

\begin{figure}
 \centering
 \includegraphics[width=0.8\textwidth]{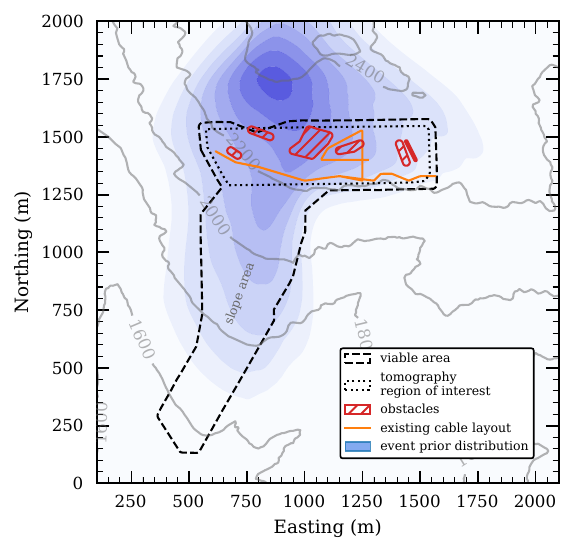}
 \caption{
  Cuolm da Vi slope instability in Central Switzerland, showing the viable area (dashed black), the shoulder area (dotted black), the obstacles (red), the existing DAS cable (orange), and the event prior distribution (blue contours, darker blue is more probable). The topography is shown in grey contour lines. The coordinates are local cartesian coordinates.
 }\label{fig:cdv_setup}
\end{figure}  

To demonstrate the design optimisation process, we use the Cuolm da Vi slope instability project  \citep{Kiers2025-AdvancingSensing} in Central Switzerland, one of the largest slow-moving slope instabilities in the Alps, as a case study. The instability spans an area of approximately 1.5~km$^2$, with an elevation difference of around 800~m (see Figure \ref{fig:cdv_setup}). Its topography and the presence of inaccessible areas provide a challenging real-world scenario for the design optimisation process. The instability is currently monitored, among other methods, by a 6.5~km long DAS cable deployed in the shoulder region of the slope \citep{Kiers2025-AdvancingSensing}. Here, we assess whether alternative layouts could provide additional information for either seismic source location or tomography.

\subsection{Design Constraints and Initial Layouts}

The viable area (the area in which the cable can be deployed) is defined by the concave hull around the locations of over 1000 autonomous nodes deployed in early summer 2022, which covered most of the shoulder and slope of the instability (see Figure \ref{fig:cdv_setup}). Topographic data was provided by the Swiss Federal Office of Topography (swisstopo).
\begin{figure}
 \centering
 \includegraphics[width=0.8\textwidth]{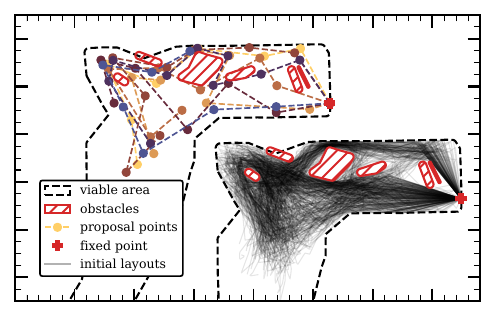}
 \caption{
  Proposal layouts and initial population for the Cuolm da Vi case study. The scale is the same as in Figure \ref{fig:cdv_setup}. The initial population has been shifted down and to the right compared to the proposal layouts for better visibility. Only a subset of 500 initial layouts is shown.
 }\label{fig:cdv_initial_layouts}
\end{figure}

For the P-wave source location scenario, we set the maximum cable length to 1700 m, which matches the length of the existing cable when excluding doubled-back segments and portions outside the viable area (see Figure \ref{fig:cdv_setup}). This constraint allows for direct comparison with the current deployment. For the Rayleigh wave tomography and multi-objective optimisation scenarios, we increase the maximum cable length to 2500 m. This extended length corresponds to the total cable length when doubled-back segments are included, providing sufficient coverage of the shoulder area to effectively perform tomography. Since the cable was not deployed with the tomography scenario in mind, a fair comparison is only possible for the source location scenario.

Unless otherwise stated, the cable takes the form of a linear (k=1) spline with 10 knots. A fixed point at (1610~m, 1330~m) ensures that the cable starts at the point where the existing cable enters the shoulder region (see Figure \ref{fig:cdv_initial_layouts}). The channel spacing is set to 10~m, which is the same as the gauge length of the existing cable for the Rayleigh wave tomography scenario, but to 100~m for the P-wave source location scenario to reduce the computational cost of the optimisation process (see Section \ref{sec:p_wave_source_location} for the justification of this choice). Within the viable area, we define several restricted areas where the cable cannot be deployed. These correspond to crevasses, boulders or rock outcrops where ground coupling is difficult to achieve, and other areas where the cable cannot be deployed (red areas in Figure \ref{fig:cdv_setup}).

The initial population is generated from manually selected proposal layouts (using an interactive plot of the study area) that cover the area of interest in a variety of distinct ways, each chosen based on heuristics and informed by expert knowledge. An intentional focus during the manual selection is placed on proposals that weave through the obstacles to ensure these paths are well represented in the initial population, as they are otherwise difficult to find during the optimisation process. If there were significantly more of these narrow paths, the generation of the initial population would need to be adapted, for example by using a unique perturbation for each proposal knot (e.g., keeping part of the proposal layout fixed, while perturbing other knots more strongly), providing a large number of proposals by hand, or changing to a different method to generate the initial population (e.g., graph-based methods, which could model the network of viable paths through the obstacles). The initial population is generated from the proposal layouts, as described in Section \ref{sec:design-optimisation}. Perturbations of $\epsilon_\mathrm{prop} = 50$~m and $\epsilon_\mathrm{knot} = 25$~m are used, along with a correlation length along the cable of 100~m. An example of a proposal layout and the derived initial population is shown in Figure \ref{fig:cdv_initial_layouts}. The example shown is specific to the P-wave source location scenario. The proposals and therefore the initial population for the other scenarios differ slightly to account for the different design criteria and cable lengths.

For the optimisation process, we use a population size of $N_\mathrm{pop} = 128$ for single-objective optimisation and $N_\mathrm{pop} = 1028$ for multi-objective optimisation for each of the 16 islands in the archipelago. The population size for multi-objective optimisation is substantially larger to ensure sufficient diversity all along the Pareto front. The population is evolved for 1000 generations with migrations occurring every 20 generations. Appendix \ref{app:optimisation_benchmarks} provides benchmarks on the trade-offs between population size, migration step-size, number of knots and spline degree.

\subsection{P-wave Source Location}\label{sec:p_wave_source_location}

First, we optimise the cable layout for P-wave source location, using the EIG as the design criterion. In addition to the design constraints above, we must define the prior distribution and the data likelihood. We assume that the event prior distribution follows the surface displacement pattern \citep{Amann2006-FAUHome}, with most potential events occurring just north of the shoulder region (see Figure \ref{fig:cdv_setup}). The horizontal components of the prior distribution are described by a normalising flow \citep{Tabak2013-FamilyAlgorithms, Dinh2014-NICEEstimation} which is fitted to the surface displacement, while the vertical component is assumed to be distributed uniformly between depths of 0 and 300~m.

Although detailed velocity models are available for parts of the instability due to an exceptionally dense monitoring campaign \citep{Kiers2025-ImagingTomography}, we use a homogeneous velocity model with a P-wave velocity of $V_\text{P} = 1500~\text{m/s}$ to better represent typical real-world scenarios where detailed subsurface information is limited. To account for a shallow low-velocity layer, we limit the maximum incidence angle to $30^\circ$, which approximates the effect that most first arrivals will be refracted and therefore have a small incidence angle.
The arrival uncertainty is determined by the picking uncertainty term ($R_\text{ref} = 1000$~m), the velocity uncertainty term ($\varsigma_\text{vel} = 0.05$), and the general observational uncertainty ($\left(\Sigma_\text{obs}\right)_{ii} = 0.01$~s) \citep{Walter2020-DistributedTerrain}. To avoid data correlations due to similar ray paths for nearby channels and for computational efficiency, we downsample the channel spacing by a factor of 10 to 100~m (taking the central channel). This spacing is large enough to avoid correlations but small enough to capture the relevant information, since a larger spacing loses only a small amount of information in the context of source location \citep{Hudson2025-TowardsNetworks}.

\begin{figure}
 \centering
 \includegraphics[width=0.8\textwidth]{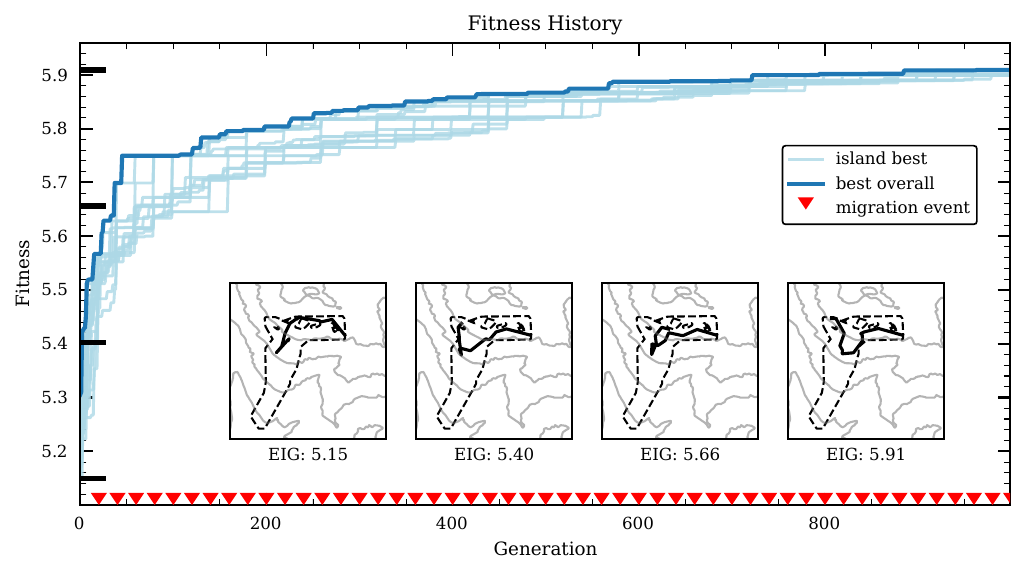}
 \caption{
  Fitness history of the EIG for the P-wave source location design optimisation. Shown are both the best layout in the archipelago (dark blue) and the best layout in each island (light blue). The four subplots at the bottom show the cable layouts at different stages of the optimisation process (indicated as black bars on the y-axis).
 }\label{fig:eig_p_fitness_history}
\end{figure}
The optimisation process begins to converge after approximately 500 generations (see Figure \ref{fig:eig_p_fitness_history}), with the best layout showing a significant improvement in the EIG compared to the initial population. The effect of migrations is clearly visible in the fitness history of individual islands, which prevents some islands from becoming trapped in suboptimal local optima and allows them to contribute to the overall convergence of the optimisation process.

\begin{figure*}
  \centering
  \includegraphics[width=1.0\textwidth]{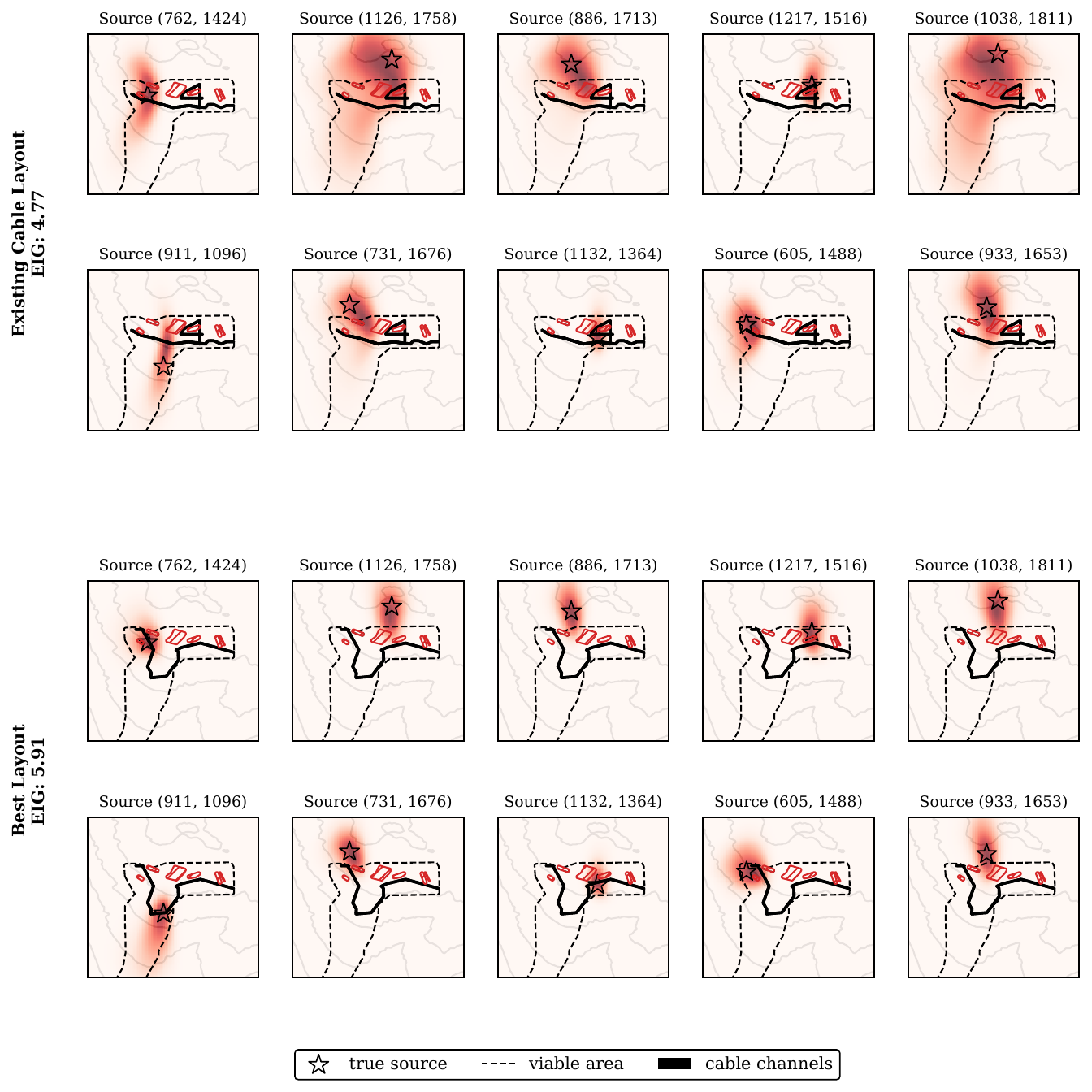}
  \caption{
  Posterior distributions of the P-wave source locations for 10 prior samples (reordered for visualisation purposes). The posterior distributions are obtained by calculating the posterior probabilities on a grid with a spacing of 20~m, and marginalising over the vertical axis. The top and bottom rows show the results for the existing and optimised cable layouts, respectively.
  }\label{fig:eig_p_posterior_plots}
\end{figure*}

While the EIG has beneficial properties for Bayesian design optimisation, it lacks a direct intuitive interpretation, which makes the benefits of optimisation difficult to appreciate. We therefore present posterior distributions for 10 prior samples in Figure~\ref{fig:eig_p_posterior_plots}, which shows directly how much better an optimised cable can resolve the source locations. If the collected data provides little information (see right column), the posterior is very similar to the prior.

\subsection{Rayleigh Wave Tomography}\label{sec:rayleigh_wave_tomography}

To demonstrate the design optimisation process for ambient noise tomography, we use Rayleigh waves, as typical sensors deployed on landslides have only a vertical component and are thus most sensitive to Rayleigh waves \citep{Feng2024-SurfaceArea, LeBreton2021-LandslideApplications}. The orientation and sensitivity of DAS cables differs inherently from those of typical seismological sensors, with substantial sensitivity to both Rayleigh and Love waves only available when aligned suitably oriented. Our tests show that optimisation for Love waves can lead to substantially different layouts, with Rayleigh waves being more sensitive to the cable layout than Love waves.

The design criterion used is D-optimality $\Sigma_D^*$ ($\lambda_\text{thr} = 10^{-4}$, $\lambda_\text{pen} = 10\log(10^{-4})$), where the sensitivity matrix is focused \citep{Curtis1999-OptimalSurveys} to contain only entries relevant to grid cells in the region of interest in the shoulder area (see Figure \ref{fig:cdv_setup}). The velocity grid is bounded by the envelope of the viable area and has a grid spacing of 20~m. The reference distance is set to 1000~m, which is conservatively low, thus implicitly optimising for a worst-case scenario in which the signal strength is just sufficient to measure group velocities across the design region at near-perfect alignment. This results in high sensitivity to inter-channel orientations.

\begin{figure*}
 \centering
 \includegraphics[width=1.0\textwidth]{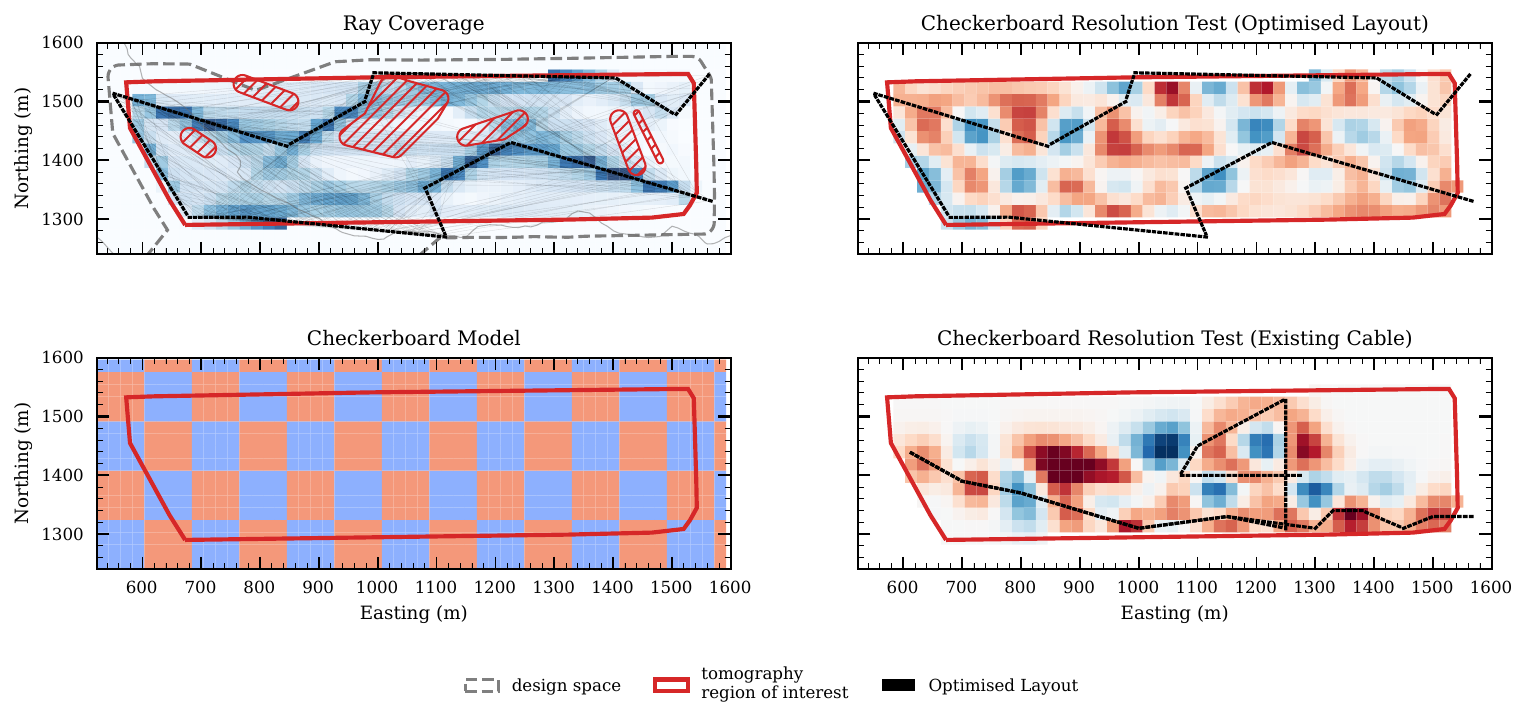}
 \caption{
  Analysis of the optimal layout for Rayleigh wave tomography. The top left plot shows the sensitivity to velocity grid cells as a heatmap (dark blue indicates high sensitivity, light blue low sensitivity), along with a subset of inter-channel ray paths (grey lines). The checkerboard models used for illustration of tomographic quality is shown bottom-left. The top right plot shows the inverted checkerboard model (red indicates high velocity, blue low velocity) that uses the sensitivity matrix of the optimal layout. The bottom right plot shows the inverted checkerboard model for the existing DAS cable layout. Each black rectangle represents a channel and its direction, which is determined by the local direction of the cable.
 }\label{fig:rayleigh_tomo}
\end{figure*}

The optimisation history follows a pattern similar to the P-wave source location case, with the best layout showing a significant improvement in D-optimality compared to the initial population. Figure \ref{fig:rayleigh_tomo} shows the optimal layout, as well as the velocity grid cell sensitivity and selected inter-channel ray paths. The figure also includes a checkerboard model used to compute synthetic data using a seismic ray-tracer \citep{Giroux2021-TtcrpyRaytracing}, and the results of inverting those data using the sensitivity matrix of both the optimal and the existing DAS cable layout.

The main pattern in the optimal layout is that the cable does not simply wrap around the shoulder region, but features notable triangular intrusions. The inter-channel ray paths show that these are necessary to enhance the angular sensitivity, since parallel channels provide no sensitivity to the velocity grid cells between them. It is important to note, however, that there are different layouts (albeit showing similar local patterns) that can perform similarly well.

\subsection{Multi-Objective Design Optimisation}

To demonstrate the multi-objective design optimisation process we optimise the cable layout for both P-wave source location and Rayleigh wave tomography simultaneously using the non-dominated sorting genetic algorithm \citep{Deb2000-ElitistNSGA-II}.

\begin{figure}
 \centering
 \includegraphics[width=1.0\textwidth]{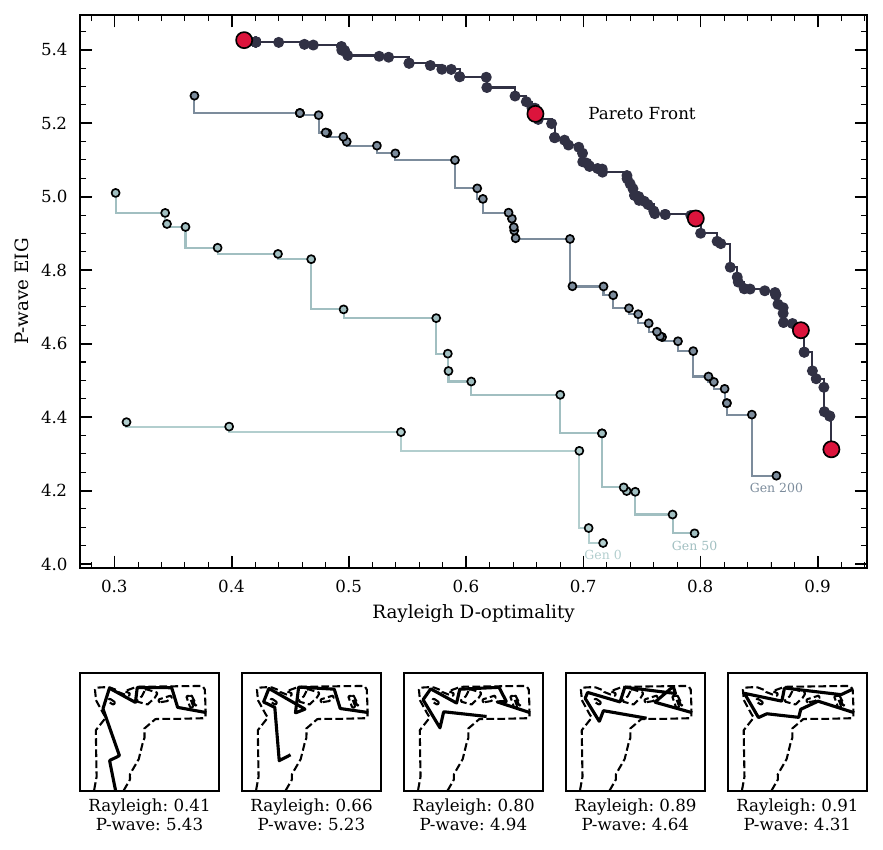}
 \caption{
  Pareto front of the archipelago population for the final (black) and previous (grey) generations of the multi-objective (P-wave source location and Rayleigh wave tomography) design optimisation process. Connecting lines are for visualisation only. Five selected layouts (red points) are shown with a solid black line in individual plots at the bottom.
 }\label{fig:pareto_front_layouts_multi}
\end{figure}

In a multi-objective optimisation process, there is no single optimal solution, but rather a set of optimal solutions known as the Pareto front. Solutions on the Pareto front represent optimal trade-offs between competing objectives, where improved performance in one criterion necessarily requires reduced performance in another. The Pareto front of the final population (and of previous generations) is shown in Figure \ref{fig:pareto_front_layouts_multi}.

We note that the end-member designs are similar in layout and slightly worse in performance compared to layouts optimised for a single objective. This is an expected outcome, as the multi-objective optimisation process aims to improve performance along the entire Pareto front, not just at the end-member designs. During the optimisation process, the number of designs on the Pareto front tends to increase as the design space is explored and new designs are found that improve performance in one or both criteria. This provides an increasingly large set of designs from which to choose.

In this specific case, the EIG criterion appears more robust to the particular layout chosen and consistently leads to good (EIG $\geq$ 4.0) performance, whereas the D-optimality criterion is more sensitive to the layout and may result in designs with low D-optimality. This can be explained by the dependence on two channel azimuths and the large model space for Rayleigh wave tomography.

\begin{figure}
 \centering
 \includegraphics[width=1.0\textwidth]{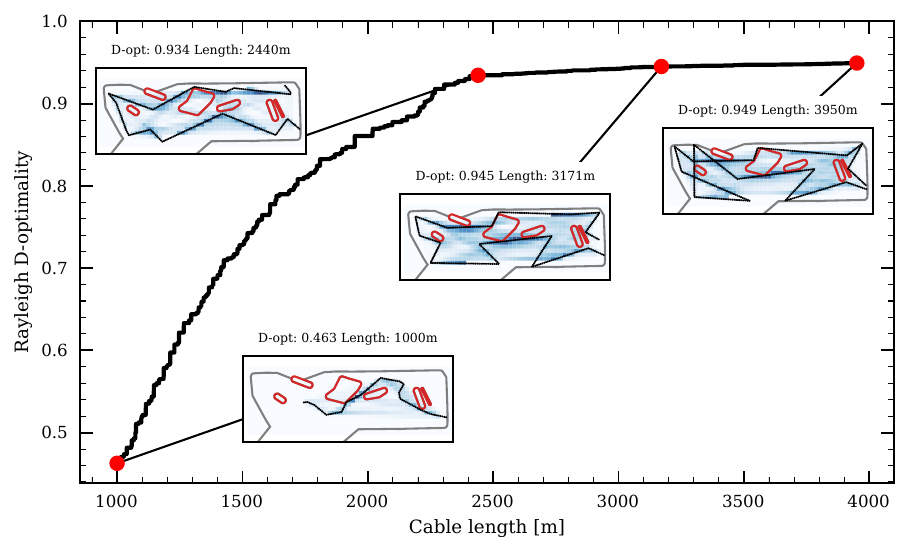}
 \caption{
  Pareto front of the multi-objective design optimisation process for cable length and Rayleigh wave tomography performance. Note that cable length is minimised rather than maximised, which flips the shape of the front compared to Figure \ref{fig:pareto_front_layouts_multi}. Four selected designs (red dots) are shown along the front, where the solid black line indicates the design and the blue heatmap shows the Rayleigh wave sensitivity.
 }\label{fig:mult_length_pareto_front_obstacles}
\end{figure}

The multi-objective optimisation process is flexible and does not require the quantities being optimised to be design performance criteria. For example, we can use the cable length as an objective to be minimised while maximising the performance of Rayleigh wave tomography, which leads to a Pareto front that shows the trade-off between cable length and D-optimality (Figure \ref{fig:mult_length_pareto_front_obstacles}). The maximum possible length of the cable has been increased to 4000~m to allow for longer layouts, and the number of knots has been increased to 16 to allow for more complex layouts.

As expected, cable length is a strong indicator of performance, with longer cables generally providing better results. The cable shapes suggest that the number of knots is more than sufficient for the shorter cables and just adequate for the longer ones. If the range of cable lengths is too wide, we therefore recommend either using a conservatively high number of knots, at the expense of convergence speed, or running several optimisation runs for different cable length ranges, each with a different number of knots.

\section{Discussion}

The DAS layout optimisation process presented here is a flexible and scalable approach to design optimisation for a wide range of DAS applications. While we have demonstrated the process only for a few specific applications, it can be adapted to any application where the design goal(s) can be expressed quantitatively. The main effort for the practitioner is to specify the prior knowledge necessary, to formulate the design criteria, define the optimisation constraints for the cable layout, and choose proposals layouts.

The choice of prior knowledge is crucial in any experimental design process \citep{Huan2024-OptimalComputations}. Introducing unreasonably strong prior knowledge may bias the design towards specific solutions, while overly weak prior knowledge may lead to designs that attempt to cover a wide range of possible scenarios without focusing on those that occur in practice. Both criteria introduced here allow prior knowledge to be incorporated in a probabilistic way, which enables prior uncertainty to be encoded to some extent. If the practitioner is uncertain about the effect of prior knowledge, they can run a multi-objective optimisation to study the effects and trade-offs of different choices. A common example of such a choice is to decide how to fix the velocity model; once chosen, this is straightforward to incorporate (at higher computational cost) and different models can lead to different designs.

All of the presented design criteria can be evaluated within seconds to minutes, depending on the exact choices and number of channels. This allows for human-in-the-loop optimisation, where the practitioner could interactively change the knot locations and other parameters and immediately see the effects on the design criteria. This is especially useful for complex scenarios where the practitioner has a good understanding of the problem but may not be able to express it fully quantitatively. The use of proposal layouts allows sequential refinement of layouts and incorporation of prior knowledge as well as lessons learned in previous optimisation runs. It also allows for sequential increases in the number of knot points and, therefore, the complexity of the design.

There are, however, limitations to the current DAS cable parameterisation and optimisation process. Placing channels at fixed intervals along the spline can result in channels that span sharp bends or kinks in the cable path. Because channel measurements represent averages over a finite gauge length, this may violate the point receiver assumption that underlies our analysis. Although this effect is unlikely to affect the optimisation significantly for applications at the scale considered here, it could be mitigated by either excluding affected channels from the analysis, subdividing them into smaller segments with appropriately increased noise levels, or integrating the sensitivity kernel along the gauge length of each channel. For splines of degree $k \geq 1$, it is important to ensure that the channel spacing is sufficiently small so that the effects of curvature remain negligible. The optimisation process presented can be extended, given the appropriate sensitivity models, to special cable geometries, such as helically wound cables \citep{Ning2018-High-resolutionDAS,Innanen2017-DeterminationWinds}, by using the appropriate sensitivity kernels. The flexibility of evolutionary algorithms would additionally allow the placement of diverse cable winding patterns as point sensors along the cable. In both cases care needs to be taken to scale the signal-to-noise ratio appropriately.

While evolutionary algorithms are flexible and scalable, they often require a large number of evaluations to converge to a solution. For computationally expensive applications, such as full waveform and high-dimensional travel time tomography \citep{Maurer2017-OptimizedImaging}, this will result in prohibitively long optimisation times, even when both the criterion evaluation and the optimisation process are parallelised. An efficient alternative to evolutionary algorithms is stochastic gradient descent (SGD) \citep[e.g.,][]{Foster2019-UnifiedExperiments,Kleinegesse2021-Gradient-basedBounds}. However, SGD requires that both the design criterion and the mapping from knots to channels are differentiable. It is also unclear how design constraints can be incorporated into the optimisation process, as SGD is not designed to handle constraints.

Another limitation of the current design optimisation process is that random crossover and mutation can struggle to respect design constraints, particularly when the viable area is small relative to the cable length or when numerous knots describe the design. This can result in many rejected layouts, which consequently slows convergence. While it remains unclear how to address this issue in general, the crossover and mutation operators could be adapted to reflect design constraints more accurately when generating new layouts.

Finally, although we have focused on the design of single DAS cables only, the design optimisation process can be straightforwardly adapted to hybrid (e.g., DAS and geophone) or multi-cable scenarios due to the inherent flexibility of evolutionary algorithms. We have omitted this from the present study, as combining data from different sensor types is a non-trivial task \citep{Hudson2025-TowardsNetworks}. If estimates for noise levels (and coupling) of different sensor types are available, the reference distance can be replaced by a source magnitude for which the hybrid design can be optimised. If the additional stations are in a low-noise environment, one could also assume that these stations record every signal (independent of distance and direction) and proceed with a reference distance only for the DAS cable as described above.

\section{Conclusion}

This study presents a flexible approach to optimising DAS cable layouts for geophysical applications. We have developed a parametric representation that uses splines to explore possible layouts efficiently while respecting physical constraints. Our adapted design criteria for source location and surface wave tomography incorporate DAS-specific characteristics such as directional sensitivity and signal decay.

Evolutionary algorithms offer robust optimisation capabilities for complex design spaces with multiple constraints, while multi-objective optimisation enables the exploration of trade-offs between competing goals. The generalised island model allows parallelisation with limited communication overhead, which makes the approach computationally feasible for realistic scenarios.

The Cuolm da Vi slope instability case study demonstrates that optimised DAS layouts significantly improve performance compared to initial designs, with different optimisation targets resulting in distinctly different cable geometries. Despite limitations in computational efficiency for very complex models, our framework supports evidence-based decision-making when planning DAS deployments for specific monitoring objectives.

\section*{Acknowledgements}
The implementations of the algorithms in this work would not have been possible without extensive use of open-source software. Not all of them have been included in the respective sections to ease readability. All the code was written in Python \citep{VanRossum2011-PythonManual}, PyTorch \citep{Paszke2019-PyTorchLibrary} was used to process probability distributions and calculate the EIG and Matplotlib \citep{Hunter2007-MatplotlibEnvironment} for plotting.

During the preparation of this work the authors used LLMs as a tool in the code development process and as a grammar and spelling checker during the writing process. After using this tool, the authors reviewed and edited the content as needed and take full responsibility for the content of the published article.

This project has received funding from the European Union’s Horizon 2020 research and innovation programme under the Marie Skłodowska-Curie grant agreement No 955515 – SPIN ITN (www.spin-itn.eu)

\section*{Data Availability}

This study only uses synthetic data. The algorithms developed in this work are available at \url{https://github.com/dominik-strutz/dased} and the code to reproduce the figures is available at \url{https://github.com/dominik-strutz/dased_paper}.

%% file: appendix.tex
\section{Optimisation Benchmarks}\label{app:optimisation_benchmarks}

In this section we show how the cable parameterisation and the main optimisation parameters affect the performance of the optimisation algorithm. For this we use the Rayleigh wave tomography setup from Section~\ref{sec:rayleigh_wave_tomography}. 

\begin{figure*}
  \centering
  \includegraphics[width=1.0\textwidth]{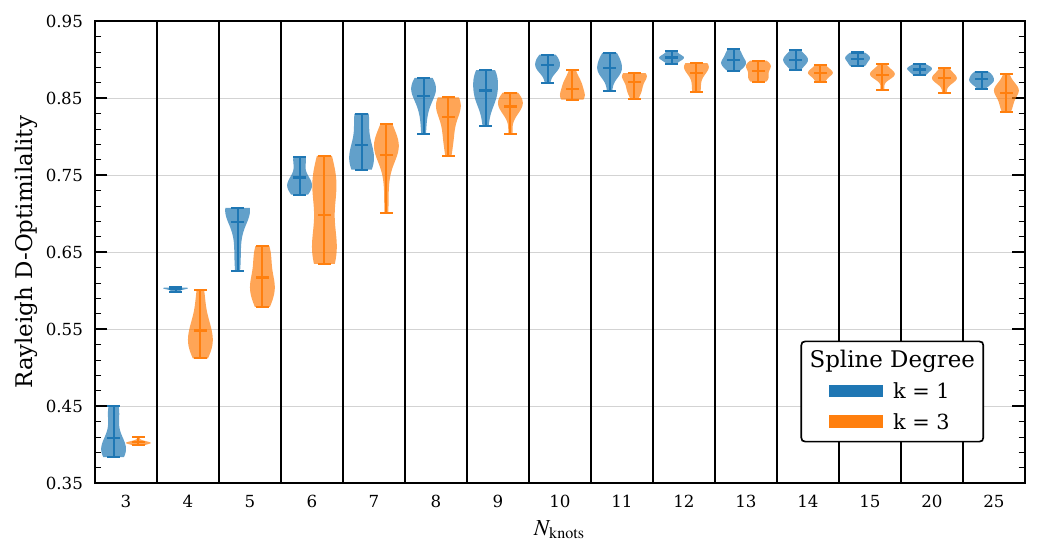}
  \caption{
  The effect of the number of knots $N_\text{knot}$ and spline degree $k$ on the fitness of the optimised cable shape. The violin plots show the distribution of fitness values for each combination of $N_\text{knot}$ and $k$, for five optimisation runs with different random seeds.
  }\label{fig:fitness_vs_N_anchors}
\end{figure*}
The number of knots $N_\text{knot}$ and spline degree $k$ determine which shapes the DAS cable can take. A larger number of knots allows for more complex shapes, but also increases the number of parameters to be optimised. The spline degree $k$ determines the continuity of the cable shape, with higher values leading to smoother shapes. In Figure~\ref{fig:fitness_vs_N_anchors} we show that an increase in $N_\text{knot}$ leads to initially better fitness, but after about ten knots the returns diminish. A linear spline ($k=1$) performs slightly better than a cubic spline ($k=3$) for the same number of knots, with the difference becoming less pronounced for larger $N_\text{knot}$. Therefore we use a linear spline with ten knots in this paper. While using more knots seems like a safer choice, convergence during optimisation is slower in the cases we tested (see results for more than fifteen knots in Figure~\ref{fig:fitness_vs_N_anchors}, which perform increasingly poorly). It is also harder to generate a good initial population for larger $N_\text{knot}$, since the random perturbation leads to a jagged cable that covers less distance than one with fewer knots. This can be alleviated by the use of correlated permutations, which artificially smooth the cable shape. 

\begin{figure*}
  \centering
  \includegraphics[width=1.0\textwidth]{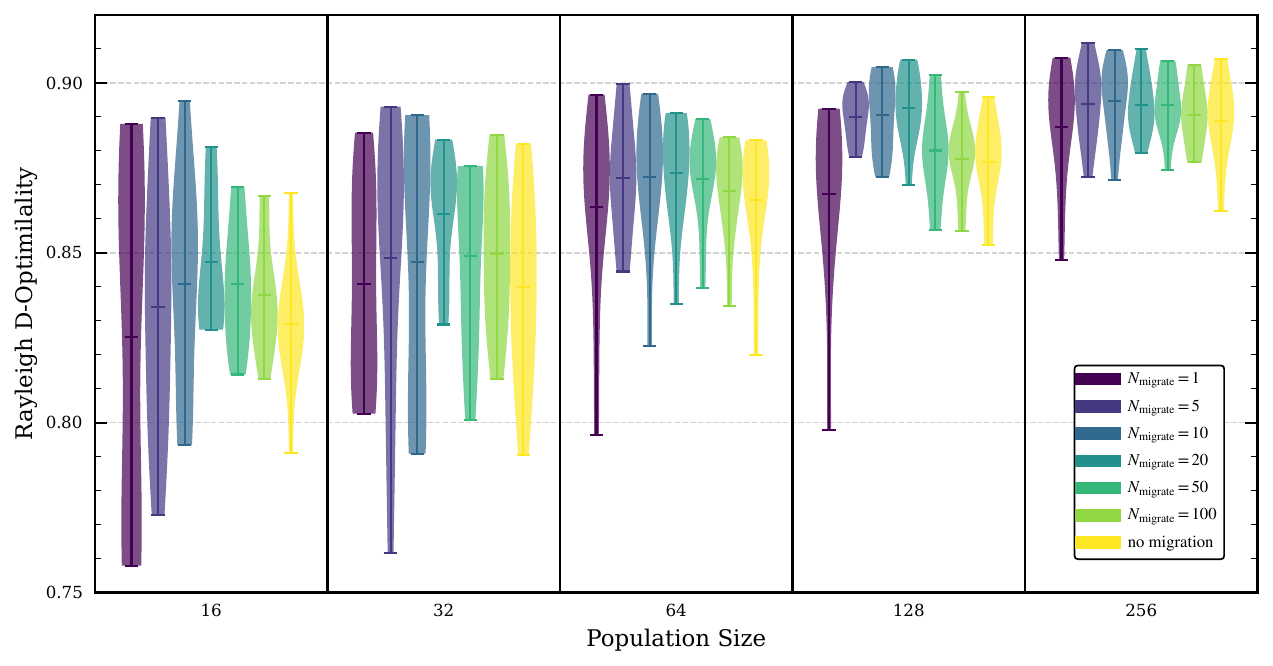}
  \caption{
  The effect of the population size $N_\text{pop}$ and migration step-size $N_\text{migrate}$ on the fitness of the optimised cable shape. The violin plots show the distribution of fitness values for each combination of $N_\text{pop}$ and $N_\text{migrate}$ for five optimisation runs with different random seeds.
  }\label{fig:fitness_vs_pop_size}
\end{figure*}
The population size $N_\text{pop}$ determines how many candidate solutions are evaluated in each generation. A larger population size allows for more exploration of the parameter space, but also increases the computational cost. The migration step-size $N_\text{migrate}$ determines how often solutions are exchanged between islands. If migrations occur too frequently, the diversity across islands is reduced, which can lead to convergence to a local optimum. If migrations occur too infrequently, the islands may not share good solutions, which can lead to slower convergence. 
In Figure~\ref{fig:fitness_vs_pop_size} we show that a large population size is the main factor in improving the fitness and making it more consistent across optimisation runs. The migration step-size has a smaller effect, with a migration every 5--20 generations leading to the best results if the population size is sufficiently large. Migration every generation leads to the worst results, which shows the benefits of separate islands compared to one large population.